\documentclass{aastex63}

\newcommand\rmi{\mathrm{i}}

\shorttitle{Thermal Bulge}
\shortauthors{Gu et al.}

\begin{document}

\title{Modeling the Thermal Bulge of A Hot Jupiter with the Two-Stream Approximation}

\correspondingauthor{Pin-Gao Gu}
\email{gu@asiaa.sinica.edu.tw}

\author{Pin-Gao Gu}
\affiliation{Institute of Astronomy \& Astrophysics, Academia Sinica\\
Taipei 10617, Taiwan}

\author{Da-Kai Peng}
\affiliation{Institute of Astronomy \& Astrophysics, Academia Sinica\\
Taipei 10617, Taiwan}
\affiliation{Department of Physics, National Taiwan University\\
Taipei 10617, Taiwan}

\author{Chien-Chang Yen}
\affiliation{Institute of Astronomy \& Astrophysics, Academia Sinica\\
Taipei 10617, Taiwan}
\affiliation{Department of Mathematics, Fu-Jen Catholic University\\
New Taipei City 24205, Taiwan}

\begin{abstract}
 We revisit the problem of thermal bulge of asynchronous hot Jupiters, using HD 209458 b as a fiducial study.  
We improve upon previous works by using a double-gray atmosphere model and interior structure from MESA as the background state, and then solve for the thermal bulge in response to the semidiurnal component of stellar insolation. 
The atmosphere model is based on the radiative transfer with Eddington's two-stream approximation. 
Two opacity cases are considered: the first introduces a greenhouse effect and the second exhibits a strong temperature inversion. We find that for the predominant thermal bulges excited by g-modes of lower orders, our results 
are qualitatively similar to the adiabatic results from \citet{AS10}. It arises because the perturbed heating due to self-absorption of thermal emissions can be significant (i.e., greenhouse effect) against Newtonian damping,
thereby leading to almost undamped thermal bulges. We also find that the contribution to the thermal bulge from the evanescent waves in the convective zone is not negligible, implying that the thermal bulge is not merely confined in the atmosphere and radiative envelope. Assuming the torque balance between the thermal and gravitational bulges, we estimate the tidal quality factor of the planet for gravitational tides to match the observed radius. The limitations of our model are also briefly discussed.
\end{abstract}



\section{Introduction} \label{sec:intro}
Since the detection of the first transiting planet HD 209458 b \citep{Charb,Henry}, a hot Jupiter of 1.36 Jupiter radii, the anomalously large size of hot Jupiters has been puzzling given their old ages. The positive correlation between their size and incident stellar flux \citep[e.g.,][]{Laughlin,Weiss,TF18} suggests that some of the intense stellar irradiation may be somewhat converted to the internal heating of the planet to slow down or even stall the gravitational contraction over time. Another piece of evidence in support of the correlation is that hot Jupiters seem to be re-inflated as their host stars evolve from the main-sequence and thus becomes more luminous \citep{Hartman,McLeod}. Several models have been proposed to explain the radius anomaly \citep[e.g., refer to the review/summary in][]{Baraffe,DJ18,TF18}.

The scenario of the quadrupole moment of the mass distribution driven by semi-diurnal thermal tides, namely a ``thermal bulge", proposed by \citet[][hereafter AS]{AS10} is one of the frequently mentioned models to account for the phenomenon \citep{LF16}.  Besides the gravitational tidal bulge, another type of tidal bulge can be produced by stellar irradiation, which differentially heats the atmospheric gas and thus redistributes the mass away from spherical symmetry. This thermally driven tidal bulge, thermal bulge in short, manifests itself as regular daily variations in atmospheric pressure on the solid surface of terrestrial planets \citep[e.g.,][]{CL,Haberle}. The effect has long been studied to potentially affect spin evolutions provided that the atmosphere is dense \citep[e.g.,][]{GS69,CL03,A19}. 

AS applied the concept to a gaseous planet and worked out the quantitative details in the case of the hot Jupiter HD 209458 b. 
In their model using a linear analysis, 
a thermal bulge is formed by thermal tidal waves related to g-modes (i.e., dynamical thermal tides instead of equilibrium tides). The torque on the thermal bulge can counteract the torque on a gravitational bulge and thus maintain a hot Jupiter in a state of an asynchronous rotation, leading to a sustainable internal tidal heating to prevent the planet from gravitational contraction due to radiative cooling from the surface. On the other hand, \citet{TF18} recently applied thermal evolution and Bayesian statistical models to infer the anomalous power as a function of incident flux that best reproduces the observed radii. They found a Gaussian shape of heating efficiency versus the equilibrium temperature $T_{eq}$  of hot Jupiters, which does not favor the thermal tide model based on the scaling law from \citet{Socrates}. It should be noted that the scaling law that is based on thermal equilibrium tides has not been tested by any systematic study in a wide range of parameter space for thermal dynamical tides, nor for the background states. As stated in 
\citet{Socrates}, the theory of dynamical thermal tides was in a state of development; this is in accordance with several of the uncertainties discussed in AS, such as the assumption of adiabatic perturbations at the base of the stellar heating layer where the radiative timescale could be comparable to the thermal forcing timescale. 

Indeed recently, \citet{AL18} included the effect of Newtonian damping in the framework of AS to determine how the non-adiabatic effect reduces the quadrupole moment in the atmosphere. The Newtonian damping/cooling is commonly referred to as a linear damping process on the temperature perturbation $T'$ over the radiative timescale $t_{th}$; i.e., the entropy exchange rate between a wave and its environment is proportional to $-T'/t_{th}$ \citep[e.g.,][]{Holton}. The method of linear energy damping provides a simple prescription to study a non-adiabatic effect with the free parameter $t_{th}$. \citet{AL18} added a pressure dependence to $t_{th}$ based on the radiative transfer simulation by \citet{Iro} to mimic a more realistic radiative timescale.
Physically, the radiative timescale $t_{th}$ depends primarily on the thermal opacity in the near-infrared over the range of typical $T_{eq}$ of hot Jupiters $\sim 1000-3000$ K. For these highly irradiated planets, CO and H$_2$O are probably primary sources of thermal opacity \citep{Madhus}. 
In this study, we restrict ourselves to an exhaustive analysis of this specific topic. We go one step further from Newtonian damping to the radiative processes in the framework of the two-stream approximation.

The two-stream approximation is a double-grey model, which considers the radiative processes based on the two opacities averaged separately over the visible and infrared bands. The model describes the incoming stellar irradiation (visible) that is absorbed in the atmosphere characterized by the visible opacity $\kappa_v$ and the outgoing thermal radiation (near-infrared) that emerges from internal radiation and reprocessed incoming stellar irradiation characterized by the thermal opacity $\kappa_{th}$. Owing to its simplicity, the approach has been widely adopted  to model observed thermal emission at infrared bands and radius evolution, as well as temperature inversions of hot Jupiters \citep[e.g.,][]{Hubeny,Hansen,Guillot,GH,Madhus}. The method may help infer the C/O ratio with chemical models in the atmosphere to constrain the actual formation site of these planets in protoplanetary disks rather than forming in situ \citep[e.g.,][]{Oberg,GM19}. 

To achieve the specific goal of non-adiabatic studies, we employ the irradiated planet suite of the open-source package MESA (Modules for Experiments in Stellar Astrophysics)\footnote{\url{http://mesa.sourceforge.net}} to establish the background states for the linear theory on thermal bulges \citep{Paxton}. MESA has been built to incorporate the atmospheric model with the two-stream approximation formulated by \citet{Guillot}. In this undertaking, we concentrate on the parameters of the hot Jupiter HD 209458 b as a fiducial study to develop a theoretical model that allows us to draw comparisons with the previous works of \citet{AS10} and \citet{AL18}. Given that $T_{eq} \sim 1500$ K, the atmosphere of HD 209458 b was postulated to be the boundary between the presence of temperature inversion on hotter planets and the absence of the inversion feature on colder planets \citep{Fortney08}. Thus far, however, the temperature inversion has only been probed in the atmosphere of a few ultra-hot Jupiters of $T_{eq}>2000$ K \citep[e.g.,][]{Madhus}. 

This paper is structured as follows. The general setup and equations are outlined in \S2. A hybrid model including linear equations pertaining to radiative transfer in the atmosphere in connection to the equations in the interior is described in \S3. In \S4, we identify the equations for quadrupole moment and torques.
The results in comparison with the previous models are presented in \S5. Finally, we summarize our results and discuss the limitations of our study in \S6.


\section{General setup and equations} \label{sec:setup}
Following the approach proposed by \citet{GO}, we consider a uniformly irradiated hot Jupiter as the background state. Hence, the background states for the atmosphere and the underlying interior structure can be described by 1-D radial profiles.  The isotropic background states of the planet with no central solid core are taken from the MESA code based on the parameters for HD 209458 b---$M_p=0.69$ $M_J$, $R_p=1.359$ $R_J$, where $M_J$ and $R_J$ are Jupiter's mass and radius respectively. The isotropic incident stellar flux is 9.93$\times 10^8$ erg/cm$^2$. Additionally, the metallicity 0.02 by mass, a default value from MESA, is adopted. Apart from an optically thick interior, we adopt MESA's gray\_irradiated
model for the atmosphere. The temperature-pressure (T-p) profile of the atmosphere in this model is implemented using the formulation by \citet{Guillot}, who employed the two-stream approximation to solve the radiative transfer equations for the atmosphere of hot Jupiters. To derive the temperature profile, a radiative equilibrium is assumed between the incoming stellar irradiation (visible) and the outgoing thermal emission (infrared). Given the ratio of the visible to infrared opacities, denoted by $\gamma = \kappa_v/\kappa_{th}$, the global averaged temperature profile---eq(49) in \citet{Guillot}---provides the top boundary condition to solve for the planet interior in the code. Apart from the stellar heating in the atmosphere, a uniform internal heating rate per unit mass is applied to the planet interior in the model of the code to maintain the planet radius in the thermal equilibrium.

Based on the background states, a linear theory can be constructed for dynamical thermal tides to produce a thermal bulge in the spherical coordinates $(r,\theta,\phi)$. 
Ignoring the Coriolis force and assuming no orbital eccentricity and spin obliquity,  we focus on the perturbations of a particular mode given by the form Re[$\xi_r(r)P_{l=2}^{m=2}(\cos \theta) \exp(i2\phi - i \sigma t)$]  (and likewise for other perturbed quantities) in response to the stellar semidiurnal irradiation, where $\xi_r$ is the radial displacement, $P_l^m$ is the associated Legendre polynomial normalized by $\sqrt{2(l+m)!/[(2l+1)(l-m)]}$, $\sigma=2(n-\Omega_p)$ is the semidiurnal thermal forcing frequency, $n$ is the orbital frequency, and $\Omega_p$ is the spin frequency of the planet. The linear equations for the vertical displacement $\xi_r(r)$, the pressure perturbation $p'(r)$, and the density perturbation $\rho'(r)$ are given by \citep[e.g.,][AS]{SW02}
\begin{eqnarray}
&& {\sigma^2 \xi_r \over p/\rho} -{dy_p \over dr} -y_p  {d\ln p \over dr}+ {d\ln p \over dr} \left( {\rho' \over \rho} \right) = 0, \label{eq:perb1} \\
&& {d\xi_r \over dr}+ \xi_r \left( {2\over r}+{d\ln \rho \over dr} \right)= - {\rho' \over \rho} +{\lambda_2 \over \sigma^2 r^2} {p\over \rho} y_p, \label{eq:perb2}\\
&& y_p-  \Gamma_1 {\rho' \over \rho} + \Gamma_1 A \xi_r = (\Gamma_3-1) {\rho T \over p} \Delta s, \label{eq:perb3}
\end{eqnarray}
where $y_{p} \equiv p'/\rho$, $A=N^2/g=(1/\Gamma_1) d\ln p/dr-d\ln \rho/dr$ is the inverse of entropy scale height, $N$ is the buoyant frequency, $\lambda_2=l(l+1)=2(2+1)=6$ is the eigenvalue of the angular part of the Laplace operator, $\Gamma_1=(\partial \ln p/\partial \ln \rho)_s$ and $\Gamma_3=1+(\partial \ln T/\partial \ln \rho)_s$ are the first and third adiabatic indices, $ -\rmi \sigma T\Delta s = \epsilon'-(\nabla \cdot {\bf F'})/\rho$, $\epsilon'$ is the perturbed heating that is thermal forcing in our problem, and ${\bf F'}$ is the perturbed energy flux that provides non-adiabatic effects to the thermal forcing. Eqns(\ref{eq:perb1}), (\ref{eq:perb2}), and (\ref{eq:perb3}) correspond to momentum, mass, and energy conservation, respectively, and are
the eqns(26), (28), and (29) in AS in the absence of stellar gravitational potential. Note that AS considered only the stellar heating $\epsilon'$ in the perturbed entropy $\Delta s$ with $\nabla \cdot {\bf F'}=0$; i.e., an adiabatic model. In the Newtonian damping model, $-\nabla \cdot {\bf F'}$ is replaced by a prescription linearly proportional to $-T'$ (please refer to \S5.5 for the details).

\section{Radiative heating and cooling terms in the linear theory} \label{sec:linear}

In this section, we describe in detail how we implement the radiative heating and cooling terms in the atmosphere and in the interior. The boundary conditions of the linear equations will be explained as well.

\subsection{The atmosphere}

The background state of the planetary atmosphere in the MESA code is obtained from the radiative equilibrium  between incoming stellar irradiation and outgoing thermal emission \citep{Guillot}:\footnote{The curvature term due to spherically coordinates is not considered for the gradient of the radiative flux $d (4\pi H)/dm$. The plane-parallel atmosphere is a reasonable assumption because the length scale of an atmosphere is significantly smaller than the planet radius.}
\begin{equation}
{dH \over dm} = \kappa_{th} (J_{th} -B)+\kappa_v J_v =0,
\end{equation}
where $J$ and $H$ (and $K$ in later sections of the paper) are the moments of radiation intensity, $m=-\rho dz \approx -\rho dr$ is the column density calculated from $r=+\infty$, $B=\sigma_B T^4/\pi$ is the source function due to the blackbody radiation, and $\sigma_B$ is the Stefan-Boltzmann constant. The subscript $th$ ($v$) indicates the quantity averaged over the thermal (visible) range of wavelengths. Physically, $J$, $4\pi H$, and $4\pi/c_{light}K$ are the radiation intensity, flux, and pressure, respectively, where $c_{light}$ is the speed of light. The perturbation of the above radiative equation $dH'/dm$, which is not necessarily zero, can be introduced to the linearized energy eq(\ref{eq:perb3}) as the source term for thermal forcing; i.e.,
\begin{equation}
-\rmi \sigma T \Delta s= {d (4\pi H') \over dm}= 4\pi \kappa_v J'_v +  4\pi \kappa_{th} \left(J'_{th} - {\sigma_B \over  \pi}4 T^3 T' \right) 
=\epsilon' -(\nabla \cdot {\bf F'})/\rho,
\label{eq:rad_perb}
\end{equation}
where any change of opacities due to the perturbations has been ignored because they are generally considered as free parameters in the model. The last term in the above equation $\propto -T'$ is analogous with the Newtonian damping with the radiative timescale $\approx t_{th}$ (see eq.\ref{eq:t_th} below), while the term associated with $J'_{th}$ is the perturbation of the absorption of thermal intensity in the near-infrared and is thus related to the greenhouse effect.  The last two terms arising from thermal radiation and equivalent to the non-adiabatic term $ -(\nabla \cdot {\bf F'})/\rho$ are not considered in AS, thereby implying that $\kappa_{th}$ and by extension $1/t_{th}$ are assumed small for simplicity in their seminal work. Strictly speaking, the common phrase ``a large thermal inertial" in the case of a large radiative timescale in the literature may be a misnomer for the two-stream approximation since the thermal trap of infrared radiation, namely the greenhouse effect, does not appear in this limit.
\citet{AL18} investigated the quadrupole moment induced by thermal tides with Newtonian damping to model a non-adiabatic effect with $t_{th}$ as a free parameter, which encompasses all the uncertainties of non-adiabatic effects. No perturbation due to the greenhouse effect was explicitly modeled in their approach. Hereafter, 
we denote 
the sum of these two thermal terms as $\epsilon'_{th}$ for a shorthand expression.

In this work, we approximate the first term on the right hand side of eq(\ref{eq:rad_perb}) to be the $m=2$ component of stellar irradiation; i.e., $\epsilon'=\epsilon'_v=4\pi \kappa_v J'_v=4\pi \kappa_v J_v c_{m=2,l=2}=\kappa_v F_* \exp (-\tau_v) c_{m=2,l=2}$, where  $\tau_v = \kappa_v p/g$ is the visible optical depth
and $c_{m=2,n=2}$ is the overlap integral describing the projection of the latitudinal heating profile onto the associated Legendre polynomial $P_{l=2}^{m=2}$\citep{GO}\footnote{\citet{GO} take into account the Coriolis effect and thus the integrand $\propto (\sin \theta) H_{m=2}(\theta)$, where $H_m$ is the Hough function \citep{CL}.}:
\begin{equation}
{2\over 3} \int_0^\pi (\sin \theta) P_2^2 \sin \theta d\theta \approx 0.76. 
\end{equation}
The above value is of the order of unity, indicating that given $m=2$, the mode $P_2^2$ is more predominately excited by the latitudinal heating than other modes with $l=4,6,8,\cdots$.\footnote{For instance, $c_{m=2,l=4}$ is 0.11. The torque associated with this mode is significantly smaller than the contribution from the quadrupole due to the smaller component of gravitational potential for $l=4$. Note that $l$ should be an even integer for $m=2$ because the stellar isolation is symmetric relative to the equator in the absence of spin obliquity.} This study is limited to this predominant mode, with $m=l=2$ as has been described in \S\ref{sec:setup} for the linear analysis. The radial dependence of the stellar heating $\epsilon'$ is the same as that adopted by AS. It should be noted that
the expression of the radiative heating is constructed assuming {\it spherical} stellar isolation once the incoming irradiation enters the atmosphere from the top boundary in order to separate variables in the linear equations \citep{GO}. Thus, it is different from $4\pi \kappa_v J_v= \kappa_v F_* \exp(-\tau_v/\mu_*)$ derived in \citet{Guillot} based on the {\it collimated}  stellar irradiation.  $\mu_*=\sin \theta \cos (\phi-\phi_*)$  with $\phi_*$ being the azimuthal location of the star. The factor $\exp(-\tau_v/\mu_*)$ is unable to be split into the separated functions of $\theta$ and $\phi$. Therefore, in the following sections, we approximate the radiative heating as isotropic rather than collimated stellar fluxes despite the inconsistency. We note that the discrepancy between the two expressions is small at low latitudes and increases at high latitudes. 

We subsequently turn our attention to the infrared terms due to the outgoing thermal emission and locate a way to solve for $J'_{th}$ in eq(\ref{eq:rad_perb}).
We perturb the radiative transfer equations for the moments of radiation intensity of outgoing thermal emission, which leads to the following linear equations for $J'_{th}$, $H'_{th}$ and $K'_{th}$:
\begin{eqnarray}
{d H'_{th} \over dm} &=& \kappa_{th} ( J'_{th} - B'),\\ 
{d K'_{th} \over dm} &=& \kappa_{th}  H'_{th},
\end{eqnarray}
where we ignore the time derivative term associated with $\partial/\partial t$ because the relaxation time for the radiative field is extremely small \citep[e.g.,][]{CG}.
Similar to the radiative transfer equation for the background state, a closure relation needs to be specified to solve the above two equations for three unknowns.
We now further assume that the first and second Eddington's coefficients that apply to the background state as proposed by \citet{Guillot}, can also apply to the perturbations; i.e., $f_{Hth}=H_{th}(0)/J_{th}(0)=H'_{th(}0)/J'_{th}(0)$ and $f_{Kth}=K_{th}/J_{th}=K'_{th}/J'_{th}$.
Hence, $K'_{th}$ can be replaced in favor of $J_{th}$, and thus the above equations to be solved can be simplified to the following closed form:
\begin{eqnarray}
{d H'_{th} \over dm} &=& \kappa_{th} ( J'_{th} - B'),\label{eq:RT1}\\ 
{d J'_{th} \over dm} &=& {\kappa_{th} \over f_{Kth}} H'_{th} \label{eq:RT2}
\end{eqnarray}
subject to the top boundary condition 
\begin{equation}
f_{Hth}=H'_{th}(0)/J'_{th}(0). 
\end{equation}

Since the gas in the atmosphere should behave like an ideal gas, $T'$ can be replaced with $p'$ and $\rho'$ using the ideal gas law; namely, $T'/T=p'/p-\rho'/\rho$. Consequently, we obtain a set of linear equations for the atmosphere from eqns(\ref{eq:perb1}), (\ref{eq:perb2}), (\ref{eq:perb3}), (\ref{eq:rad_perb}), (\ref{eq:RT1}), and  (\ref{eq:RT2}) for $y_p$, $\xi_r$, $H'_{th}$, and $J'_{th}$. The specific form of the equations are listed in Appendix A. 
Note that to make the ODEs in Appendix A concise and more physically clear,
we have defined the radiative timescale
\begin{equation}
t'_{th} \equiv (2\pi) t_{th} =(2\pi)\left(  {16 \rho \kappa_{th} \sigma_B T^4 \over p} {\Gamma_3-1 \over \Gamma_1}\right)^{-1} 
\approx 0.5 {(\Gamma_1/1.37) \over 0.37/(\Gamma_3-1)} \,{\rm days} \left( {\kappa_{th} \over 0.006\, {\rm cm^2/g} }\right)^{-1} \left( {T \over 1500\,{\rm K}} \right)^{-3} \left( {\mu \over 2.36\,{\rm g/mol} }\right)^{-3} \label{eq:t_th}
\end{equation}
Using the thermal optical depth $\tau_{th}=\kappa_{th} p/g$ \citep{Guillot}, the expression of $t_{th}$ can be recast to $-c_p p/(16 \sigma_B T^3 \rho_s g \tau)$ that is similar to the radiative timescale used in \citet{GS02} with  $\tau_{th} \sim 1$, $c_p$ being the specific heat at constant pressure, and $\rho_s$ being the negative of the volume expansion coefficient at constant pressure given by $c_p (\partial \ln \rho/\partial s)_p=-1$. When $t_{th} \sigma \gg 1$---i.e., when the non-adiabatic effects are irrelevant during the period of the thermal forcing---the coefficients in the above equations become small except for $f_1$, $g_1$, $f_2$, $g_2$, $\alpha_5$ and $\alpha_{12}$. Consequently, the set of ODEs can be reduced to the ODEs in the adiabatic limit  (i.e., eqns(\ref{eq:perb1})-(\ref{eq:perb3}) without the non-adiabatic term in the energy equation). In contrast, when $t_{th} \sigma \ll 1$, the source term in eq(\ref{eq:S1_R}) still remains small compared to the limit for $t_{th} \sigma \gg 1$ and the other source term in eq(\ref{eq:S1_I}) is reduced by a factor of $[1-(1/t_{th}\sigma)^2/(1+1/(t_{th} \sigma)^2] \approx (t_{th} \sigma)^2$.   As estimated in eq(\ref{eq:t_th}), the typical radiative timescale $t'_{th}$ is approximately 0.5 days 
in comparison with the forcing period of a few days associated with the g modes of lower order  such that $t_{th} \sigma \gtrsim 1$.

At the top boundary, in addition to the upper boundary condition $f_{Hth}=H'_{th}(0)/J'_{th}(0)$ from Eddington's approximation, we follow AS and adopt an outgoing-wave or free-surface boundary condition. Under the the WKB approximation (i.e. $H_p \equiv (d\ln p/dr)^{-1} \ll r$) and for an isothermal atmosphere, the outgoing-wave condition reads 
\begin{equation}
{d\over dr} \left( {p' \over \rho} \right)= b \left( {p' \over \rho} \right),
\end{equation}
where $b=1/(2H_p) \pm \rmi [k^2_\perp N^2/\sigma^2 - k^2_\perp + \sigma^2/c^2 - 1/(4H^2) ]^{1/2}$, $k^2_\perp=2(2+1)/r^2$ for $l=2$, $H_p=p/(\rho g)$, and $c^2=\Gamma_1 p/\rho$. When the expression in the square root of Im[$b$] is positive, it admits a wave solution and the sign of Im[$b$] should be selected such that the wave energy flux is positive (i.e. outgoing) across the boundary. When the expression in the square root of Im[$b$] is negative, the sign of Im[$b$] is selected such that an evanescent wave occurs beyond the top boundary. For a free surface, the Lagrangian pressure perturbation vanishes; namely, $p'/\rho=g \xi_r$ is set at the top boundary. 

\subsection{The interior}

In the interior underlying the atmosphere, the gas is optically thick and we adopt the Rosseland approximation to model the perturbation of the radiative diffusion. The perturbation of the imposed internal heating to maintain the radius is neglected since the internal tidal heating is not modeled in the present work. Following \citet{SP83} and \citet{SW02}, eq(\ref{eq:perb3}) is then written as\footnote{Although the stellar irradiative heating $\epsilon'_v$ is ignored in the equation for the planet interior, we observe that the solutions are almost the same when the heating is included.}
\begin{equation}
y_p-  \Gamma_1 {\rho' \over \rho} + \Gamma_1 A \xi_r = \rmi \eta \left[ {1\over Fr^2} {d(r^2 F'_r) \over dr} - \lambda_2 {\Lambda \over r^2} \left( {T' \over T} \right) \right], \label{eq:pert3}
\end{equation}
along with
\begin{equation}
{F'_r \over F}=\Lambda {d \over dr} \left( {T'\over T} \right) + (4-\kappa_T){T' \over T} - (1+ \kappa_\rho) {\rho' \over \rho}, \label{eq:pert4}
\end{equation}
where 
\begin{eqnarray}
\Lambda &=& \left( {d \ln T \over dr} \right)^{-1}, \label{eq:L}\\
F&=& -{4ac T^3 \over 3 \kappa \rho}  {d T \over dr}= -{16 \sigma_B T^4 \over 3 \kappa \rho \Lambda} \Longrightarrow F\Lambda = -{16 \sigma_B T^4 \over 3 \kappa \rho}. \label{eq:FL}\\
\eta &=& -(\Gamma_3 -1 ) {F \over \sigma p},\\
\kappa_T&=&\left( {\partial \ln \kappa \over \partial \ln T} \right)_{\rho}, \qquad \kappa_\rho=\left( {\partial \ln \kappa \over \partial \ln \rho} \right)_T. \label{eq:dkappa}
\end{eqnarray}
The term $\eta$, which consists of $\sigma$ and the radiative timescale $\propto p/F$, describes a characteristic length scale for radiative diffusion. It is expected to be significantly smaller than the planet radius deep in the interior but might not be so near the top boundary. 
We conclude with the linear equations for $y_p$, $\xi_r$, $F'_r$, and $y_T\equiv T'/T$  for the planet interior that are governed by eqns(\ref{eq:perb1}), (\ref{eq:perb2}), (\ref{eq:pert3}), and (\ref{eq:pert4}). The explicit forms are listed in Appendix B. The bottom boundary conditions $\xi_r=F'_r=0$ at the planet center are applied due to the spherical symmetry of the problem.

\begin{figure}
\plotone{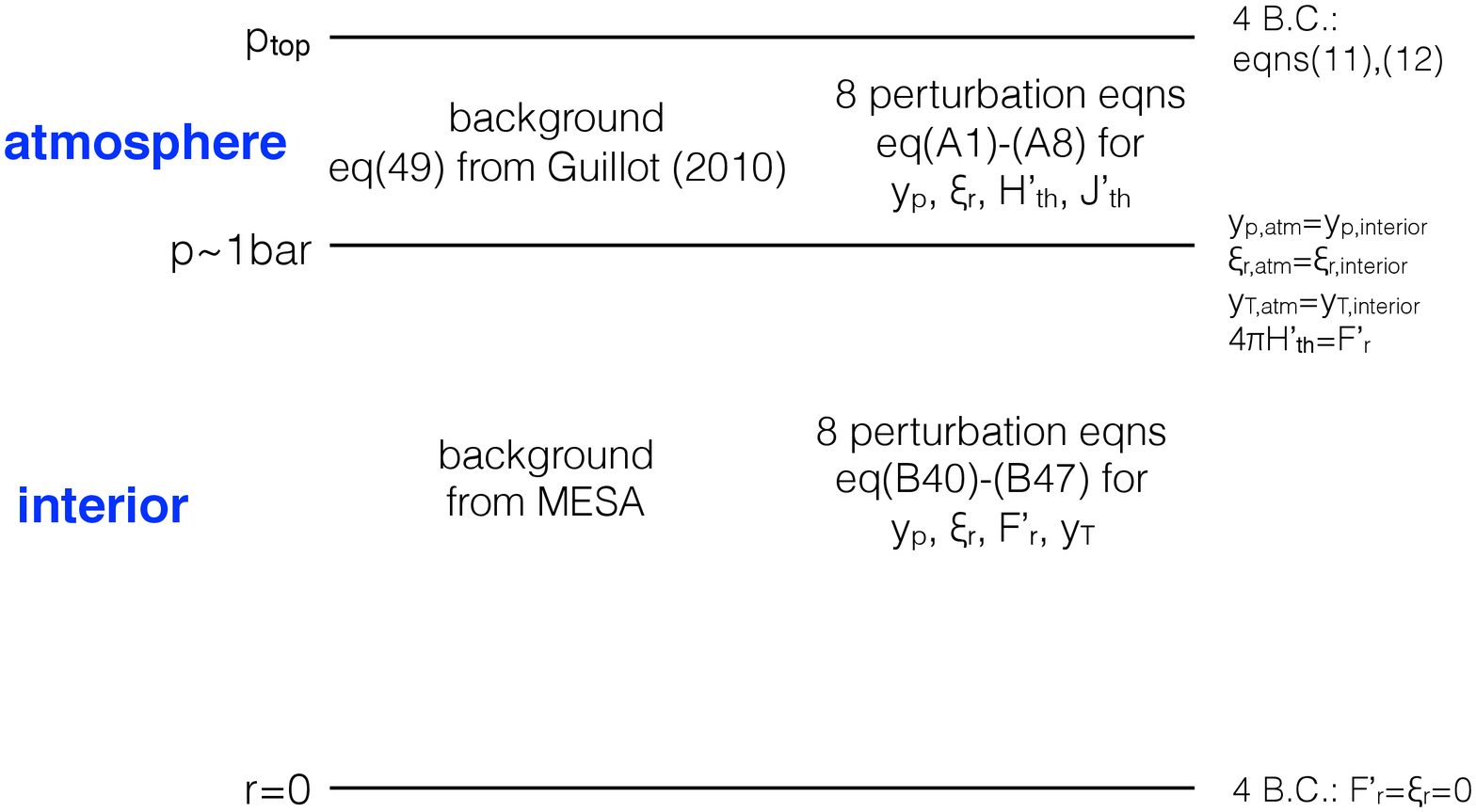}
\caption{Summary of our hybrid model for thermal tides.}
\label{fig0}
\end{figure}

\subsection{The hybrid model}

Finally, we gather one set of linear equations for the atmosphere and another set for the planet interior.  Together, these provide a ``hybrid" model to describe the linear response of the entire planet to the semi-diurnal component of stellar thermal forcing $\epsilon'_v$ in the atmosphere. We employ the finite difference to the 2nd order accuracy to discretize the linear equations from the atmosphere down to the planet interior. The procedure gives the combined linear equations with the coefficients in a form of a sparse matrix as done by AS. While solving the entire discretized equations by inverting the sparse matrix, the perturbed variables should be connected at the interface between the atmosphere and interior. Since the base of the atmosphere lies at the top boundary of the computation domain of the planet in the MESA code, the location of the interface needs to be properly selected by hand. It should be deep enough for the majority of stellar heating $\rho \epsilon'_v$ to be included in the atmosphere. Otherwise, most of the stellar heating would be excluded from the code. Furthermore, the location of the interface should be shallow enough that the optical depth for the thermal emission is not too large to be optically thick. Clearly, a plausible choice of the location is to place the interface at the base of the stellar heating layer. For the visible and thermal opacities that we use, the interface generally falls around the pressure $p\sim 1$ bar (see Fig~\ref{fig1} later in the paper).

In summary, the background equilibrium state is provided from MESA with the location of the top boundary set to be around the $p \sim 1$ bar, above which the T-p profile is governed by eq(49) in \citet{Guillot} under the assumption of radiative equilibrium\footnote{As explained in \S2, for the gray\_irradiated model in the MESA code, the $T-p$ profile from eq(49) of \citet{Guillot} provides the top boundary condition to produce the interior structure of the planet. Therefore there is no direct output of the $T-p$ profile for the atmosphere from MESA. The background state has to be reconstructed using eq(49) of \citet{Guillot}.}. For the linear perturbations for thermal tides, we solve the linear equations from eq(\ref{eq:D_first}) to eq(\ref{eq:D_last}) in the atmosphere and eq(\ref{eq:MESAf}) to eq(\ref{eq:MESAl}) in the planet interior. Two bottom boundary conditions exist---$\xi_r=F'_r=0$ at $r=0$. The other two conditions at the top of the atmosphere ($r=r_{top}$) are the outgoing-wave/free-surface condition and the Eddington approximation. Given the ODEs and boundary conditions, the perturbed variables $y_T$, $y_p$, $\xi_r$, and $F'_r=4\pi H'$ should be continuous at the interface between the atmosphere and interior. The expressions of $y_T$ in the atmosphere, which do not appear explicitly in the linear equations, are given by eqns~(\ref{eq:T'_R}) and (\ref{eq:T'_I}) to use for the connection.
The summary of our hybrid model is illustrated in Fig.~\ref{fig0}. Following proper discretization, the linear equations throughout the entire planet can be solved by the inverse of a sparse matrix that consists of the coefficients in the discretized linear equations.

\section{Quadrupole moment, torques, and planet inflation}

After the density perturbation is computed, the imaginary part of quadrupole moment of  the mass distribution Im[$Q_{22}$] and the resulting time-averaged torque can be computed by (e.g., AS)
\begin{equation}
T=4\left(  {3\pi \over 10} \right)^{1/2}  n^2 {\rm Im}[Q_{22}]=
4\left(  {3\pi \over 10} \right)^{1/2}  n^2 \int_0^{r_{top}} {\rm Im}[\rho'(r)] r^4 dr, \label{eq:torque}
\end{equation}
where the quadrupole moment $Q_{22} $ is given by $\int_0^{r_{top}} \rho'(r) r^4 dr$.
In practice, the above equation can be recast to a form in terms of $\xi_r$ and $\rho'$ to produce a more accurate result. Please refer to eq(A3) in AS and eq(D3) in \citet{AL18}. We employ their equations to compute the torque.
The phase angle of the quadrupole moment relative to the stellar irradiation is given by $\Delta \phi = (1/2)\arctan$ [Im($Q_{22})/$Re($Q_{22})$]. In this study, using the wave form $\propto \exp(2\phi-\sigma t)$, whenever $\sigma<0$ (i.e., retrograde thermal forcing), $\Delta \phi >0$ is required for the thermal bulge to lead the stellar irradiation; this results in a positive torque (i.e., $T>0$) against the negative torque due to the gravitational tidal bulge. On the other hand, when $\sigma>0$ (i.e. prograde thermal forcing), $\Delta \phi <0$ is required for the thermal bulge to lead the stellar irradiation and thus generates a negative torque (i.e., $T<0$) against the positive torque due to the gravitational tidal bulge\footnote{This is the case considered in AS. Note that in the absence of the Coriolis effect, the quadrupole moment and consequently the torque do not change magnitudes but merely flip the sign when $\sigma$ changes the sign.}. Since the exact equilibrium tide occurs when $\sigma=0$, AS showed that there exists an analytic solution for the thermal quadrupole moment in the limit of small $\sigma$:
\begin{equation}
Q_{22}=4\int_0^{r_{top}} dr \rho r^4 {\sigma^2 \over N^2}{\Gamma_3 -1 \over \Gamma_1} {\rho T \over p} \Delta S. \label{eq:analytic}
\end{equation}
In the model with Newtonian damping, \citet{AL18} showed that the entropy perturbation $\Delta S$ is reduced to the stellar heating term $\epsilon'_v$ in the above equation. 
The same result also applies to our hybrid model in this limit because the behavior of the non-adiabatic terms including the greenhouse effect (i.e., eq~\ref{eq:rad_perb}) are governed by the same parameter $\sigma t_{th}$.

In the dynamical equilibrium, the gravitational torque in eq(\ref{eq:torque}) due to the thermal bulge should be balanced by that due to the gravitational bulge given by \citep[e.g.,][]{Zahn,Gu03}
\begin{equation}
T_{grav}={9 G M_*^2 R_p^5 \over 2 a^6 Q_p'} \left( {\Omega_p \over n} -1 \right)
\approx 2.8\times 10^{30} \left( {M_* \over M_\odot} \right)^2 \left( {R_p \over 1.36R_J} \right)^5
\left( {a \over 0.047{\rm AU}} \right)^{-6} \left( {Q'_p \over 10^8} \right)^{-1} \left( {\Omega_p \over n} -1 \right)\, {\rm dyne\,cm},\label{eq:T_grav}
\end{equation}
where we adopt the prescription for the equilibrium tides with a constant time lag that is characterized by the tidal quality factor $Q_p'$. Consequently, given a value of $Q_p'$, a set of solutions for $\Omega_p$ and thereby for the forcing period can be obtained from the torque equilibrium. Besides the torque equilibrium, the thermal equilibrium provides another constraint to obtain a unique solution; i.e., the tidal heating $\sigma T_{grav}$ is balanced by the radiative cooling to maintain the observed radius of an inflated planet. With an interior model from the MESA code, the value of the free parameter $Q_p'$ can be obtained from these two equilibria.

\section{Results} \label{sec:result}
\subsection{Background state in two opacity cases}
\begin{figure}
\plottwo{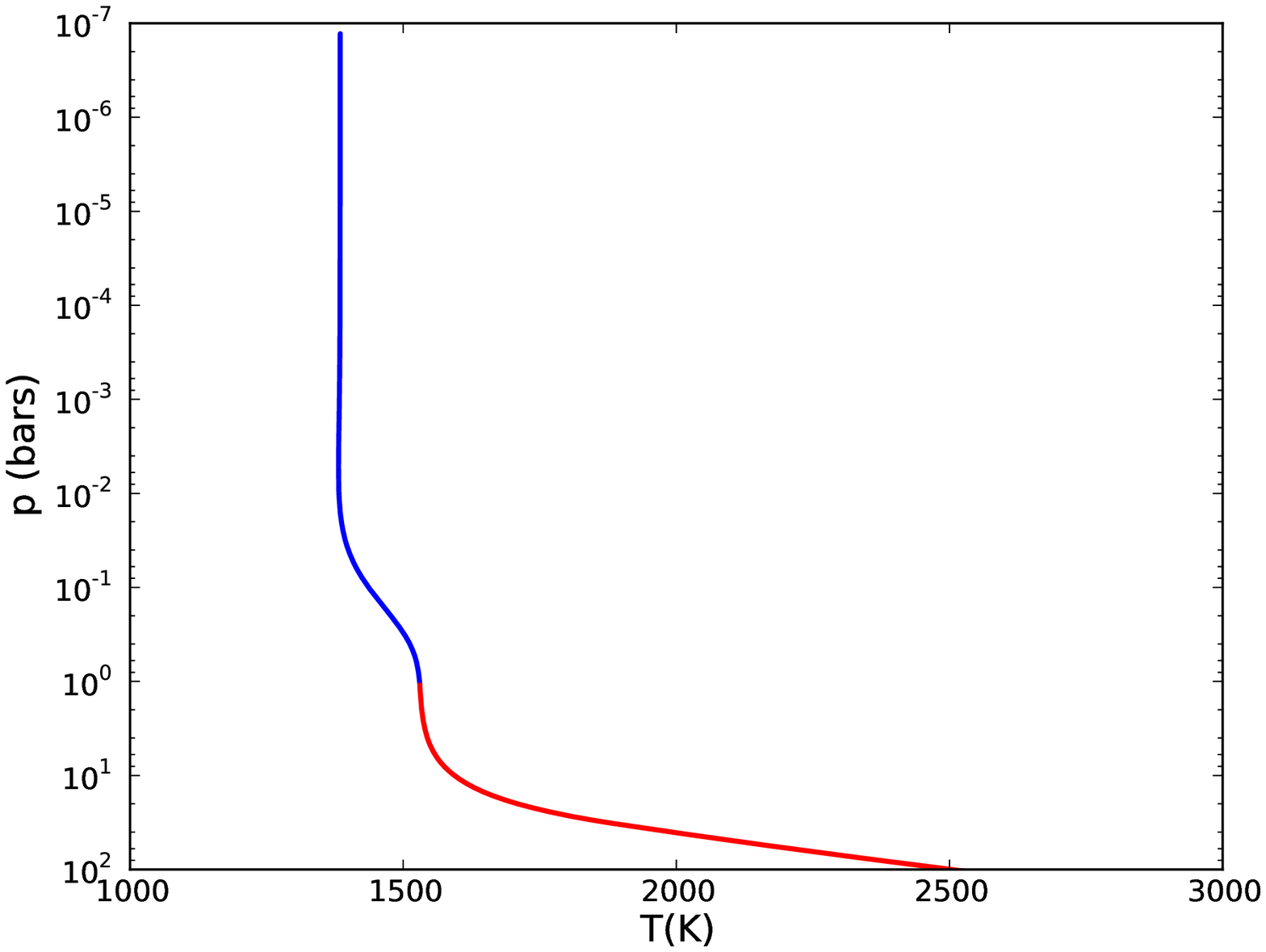}{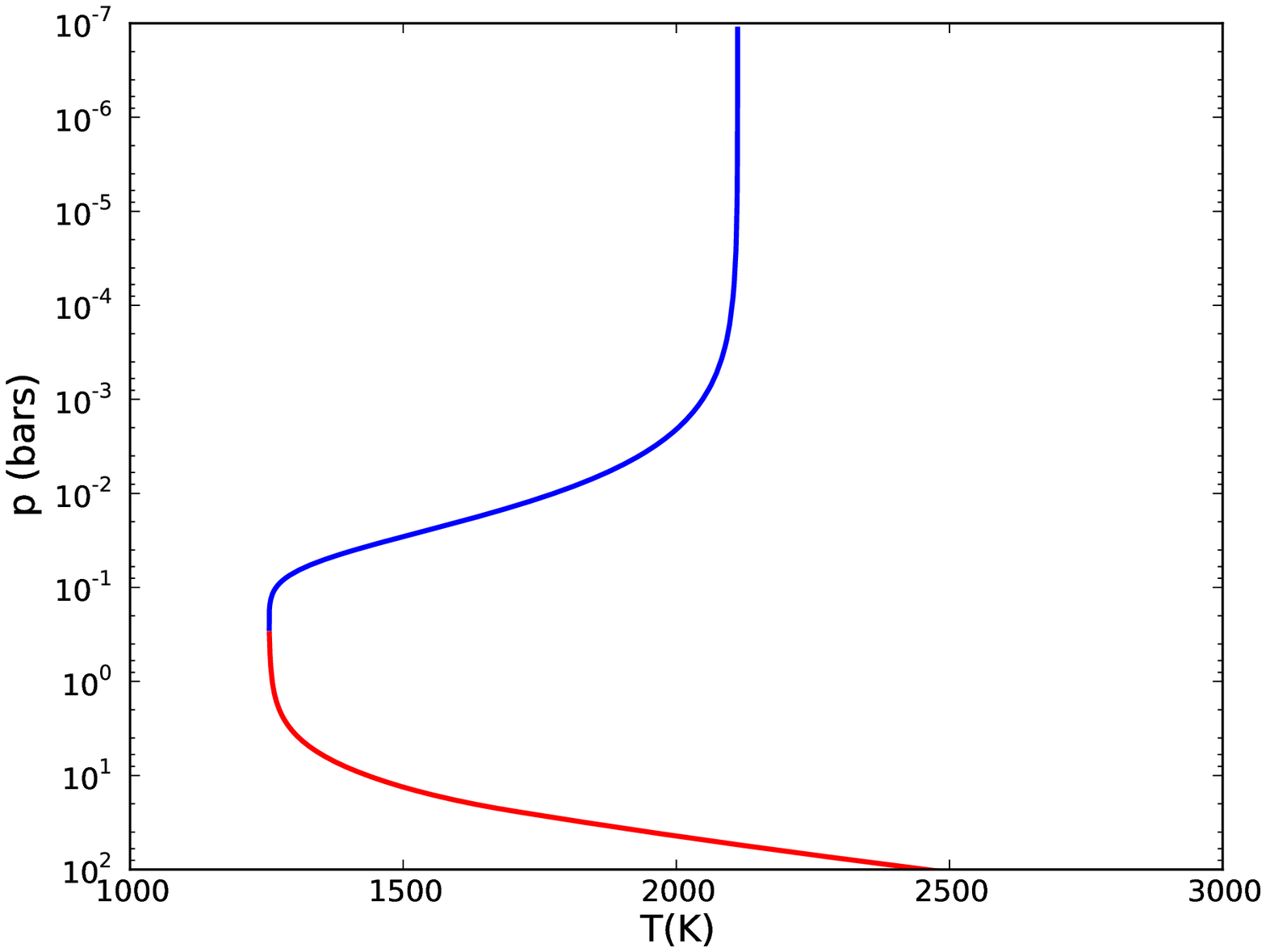}
\caption{T-p profile of the planet down to $p \sim 100$ bars for the two opacity cases: $\kappa_v,\kappa_{th}=$0.004,0.006 cm$^2$/g (left panel, $\gamma \approx 0.67$) and 0.03, 0.0037 cm$^2$/g (right panel, $\gamma \approx 8$). The profile of the atmosphere is shown in blue and that of the interior is plotted in red. The interface between the atmosphere and interior is set to fall at the base of the stellar heating layer (refer to the middle panels of Figure~\ref{fig3}).} 
\label{fig1}
\end{figure}

Since the T-p background profile of an atmosphere depends on the opacity ratio $\gamma$ , we consider two opacity cases in this study. These are $(\kappa_v,\kappa_{th})=$(0.004,0.006) and (0.03,0.0037) in units of cm$^2$/g, which correspond to $\gamma \approx $ 0.67 and 8, respectively. The former $\gamma$ value is similar to the one used by \citet{Guillot} in his fiducial model for HD 209458 b, whereas the latter $\gamma$ value is selected to be large for comparison. The left panel of Figure~\ref{fig1} presents the case of $\gamma=0.67$ where the stellar irradiation penetrates deep down to the 1-bar level and heats the atmosphere by the reprocessed stellar isolation in the near-infrared; in other words, the greenhouse effect is effective. In contrast, the right panel shows the case of $\gamma \approx 8$ in which the stellar irradiation primarily heats the upper atmosphere at 0.1$-$0.01 bar, leading to the temperature inversion at high altitudes. The T-p profile of the planet interior (plotted in red) is connected to that of the planetary atmosphere (plotted in blue) at the base of the stellar heating layer, which falls around the pressure level of $\sim 1$ bar as indicated by the solid curve for $\rho \epsilon'_v$ in the middle panels of Figure~\ref{fig3}. The connection lies slightly deeper in the case of $\gamma \approx 0.67$ due to the deeper stellar heating

To maintain the planet radius in the thermal equilibrium, the internal heating imposed in the MESA code is $5.84\times 10^{26}$ erg/s for $\gamma=0.67$ and $7.15\times 10^{26}$ erg/s for $\gamma=8$. The larger internal heating for the larger $\gamma$ is expected since the stellar heating has penetrated less into the atmosphere due to the larger $\kappa_v$ \citep{Guillot}. The corresponding heating efficiency defined as the ratio of the internal heating to the incident irradiation energy is approximately 0.2\%, which is about one order of magnitude smaller than the heating efficiency for the hot Jupiters of $T_{eq} \approx 1500$ K in the statistical analysis conducted by \citet[][see their fig 11]{TF18}. In that work, the authors used the mass-metallicity relation derived for warm Jupiters from \citet{Thorn16} as a prior for their analysis. The relation yields about a metal abundance of 21\% by mass for HD 209458 b. It is probable that the considerably smaller mean molecular weight due to the lower metallicity that we adopt ($Z=0.02$) for the planet is one of the primary causes of the significantly lower heating efficiency \citep[e.g.,][]{Gu04,Baraffe08}.

Based on the background state, we present the results from our linear analysis in the following subsections.

\subsection{Effects of boundary conditions and greenhouse}

\begin{figure}
\plottwo{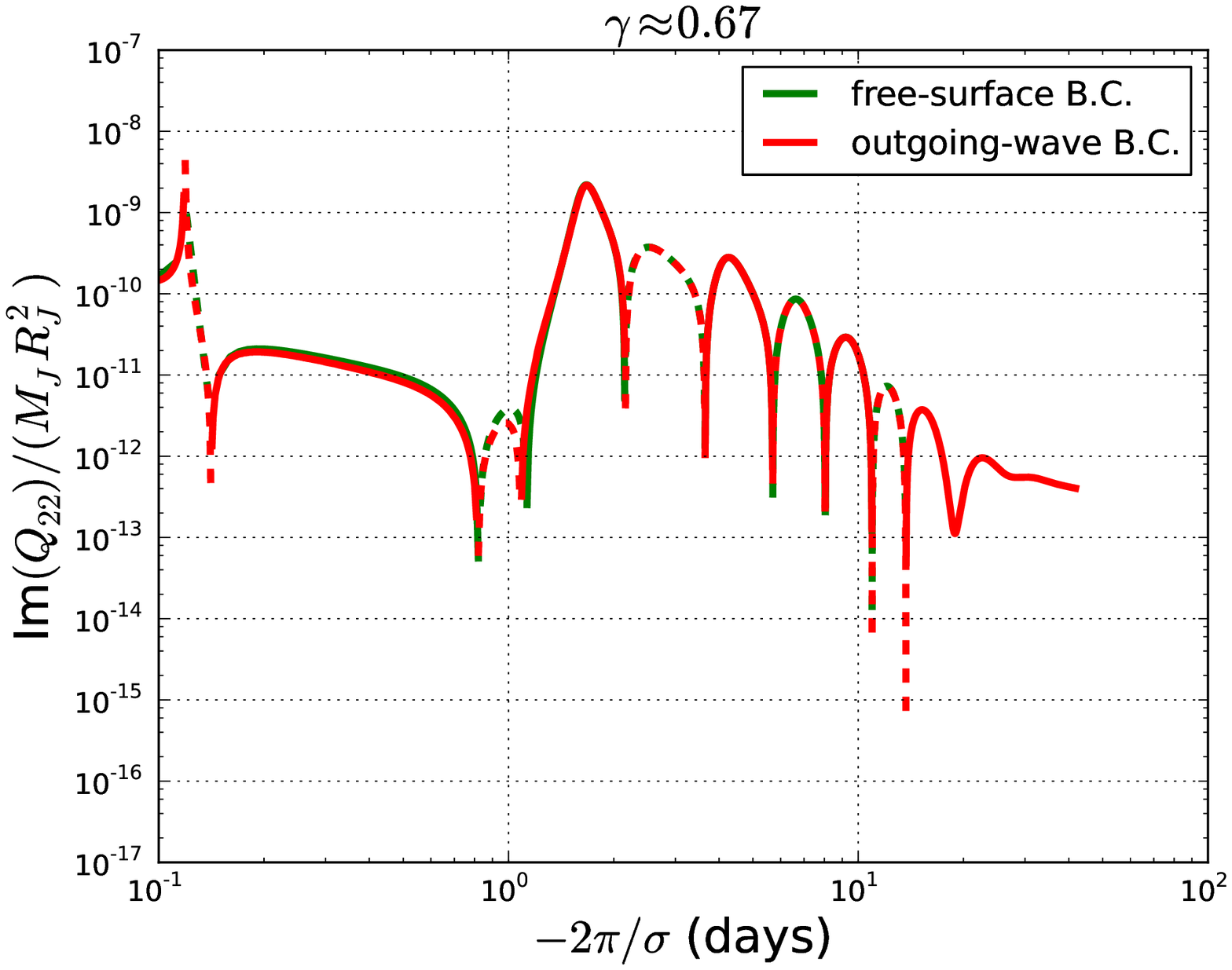}{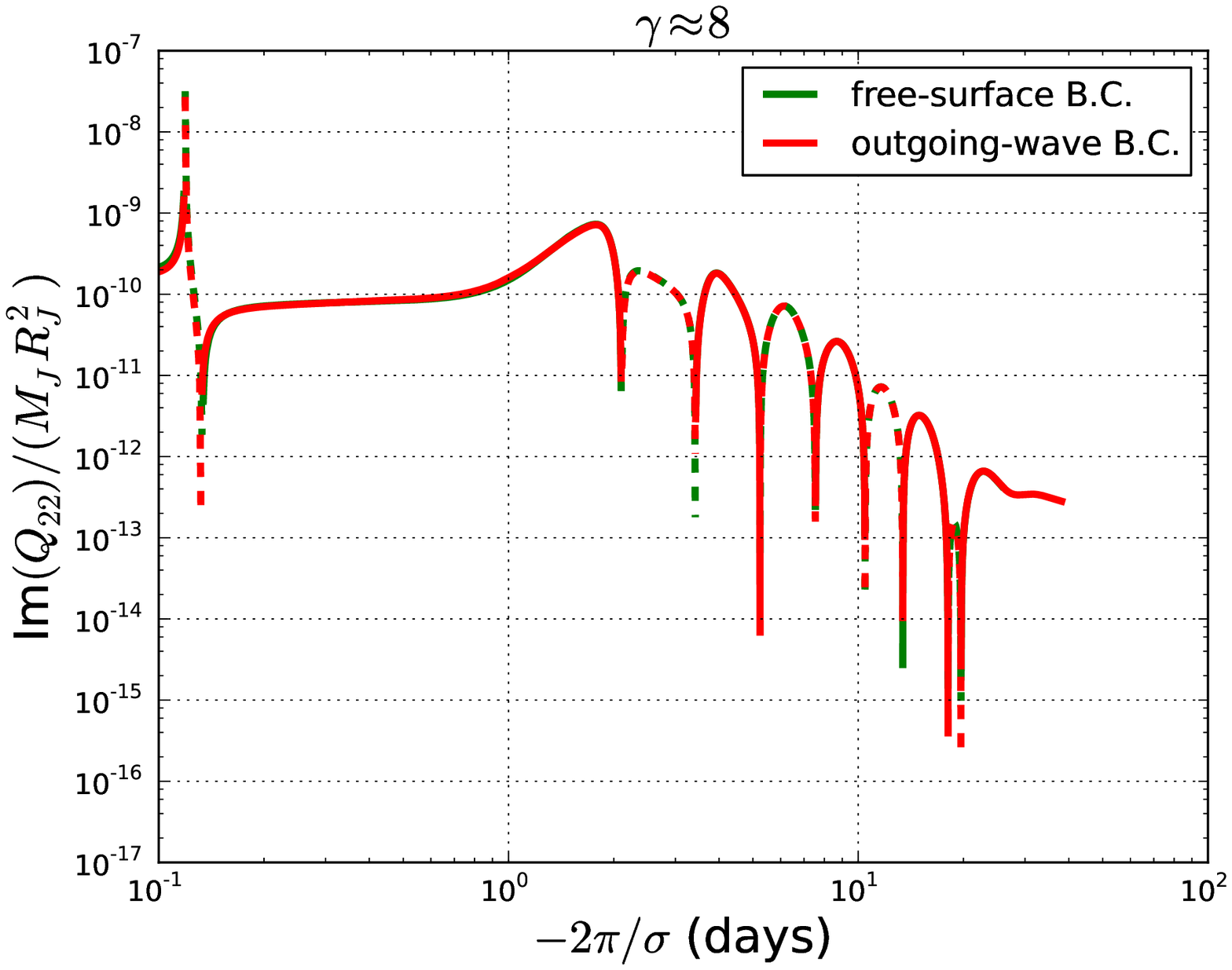}
\caption{The imaginary part of the quadruple moment Im[$Q_{22}$] as a function of forcing period with two different top boundary conditions. Two opacity cases are presented: $\gamma \approx 0.67$ (left panel) and 8 (right panel).}
\label{fig1.5}
\end{figure}

Figure~\ref{fig1.5} shows Im[$Q_{22}$] from our model as a function of the minus of the forcing period $-\sigma$ (i.e. quadrupole moment spectrum) with two different top boundary conditions: free surface and outgoing wave. The left and right panels of the figure correspond with the left and right panels of Figure~\ref{fig1} to present the results in the two opacity cases. The solid curves correspond to positive values of $\Delta \phi$ and are hence associated with the torques against synchronization; in contrast, the dashed curves  correspond to the negative values of $\Delta \phi$ and are related to the torques for synchronization alongside the torques on gravitational bulges.
We observe no noticeable distinction between the results subject to the two different top boundary conditions. The outgoing-wave condition based on the WKB approximation is less appropriate for the modes with longer radial wavelengths in the limit of smaller forcing period of the spectrum \citep[e.g.,][]{AL18}. For the background states in both opacity cases, nevertheless, the outgoing-wave boundary condition becomes the condition for the evanescent/reflected waves when $|2\pi/\sigma|<$ 1.2 days because the term $k^2_\perp N^2/\sigma^2$ is smaller than $1/4H_p^2$ in the expression of Im[$b$]. For the dissipationless thermal tides with a free surface on the top, AS showed that the g-mode spectrum exhibits sharp and dense resonant peaks, As is evident, the non-adiabatic terms smear out the  resonant peaks in the spectrum and thereby make the results insensitive to the boundary conditions. Therefore, we focus only on the results with the outgoing-wave boundary condition in the rest of the paper.

\begin{figure}
\plottwo{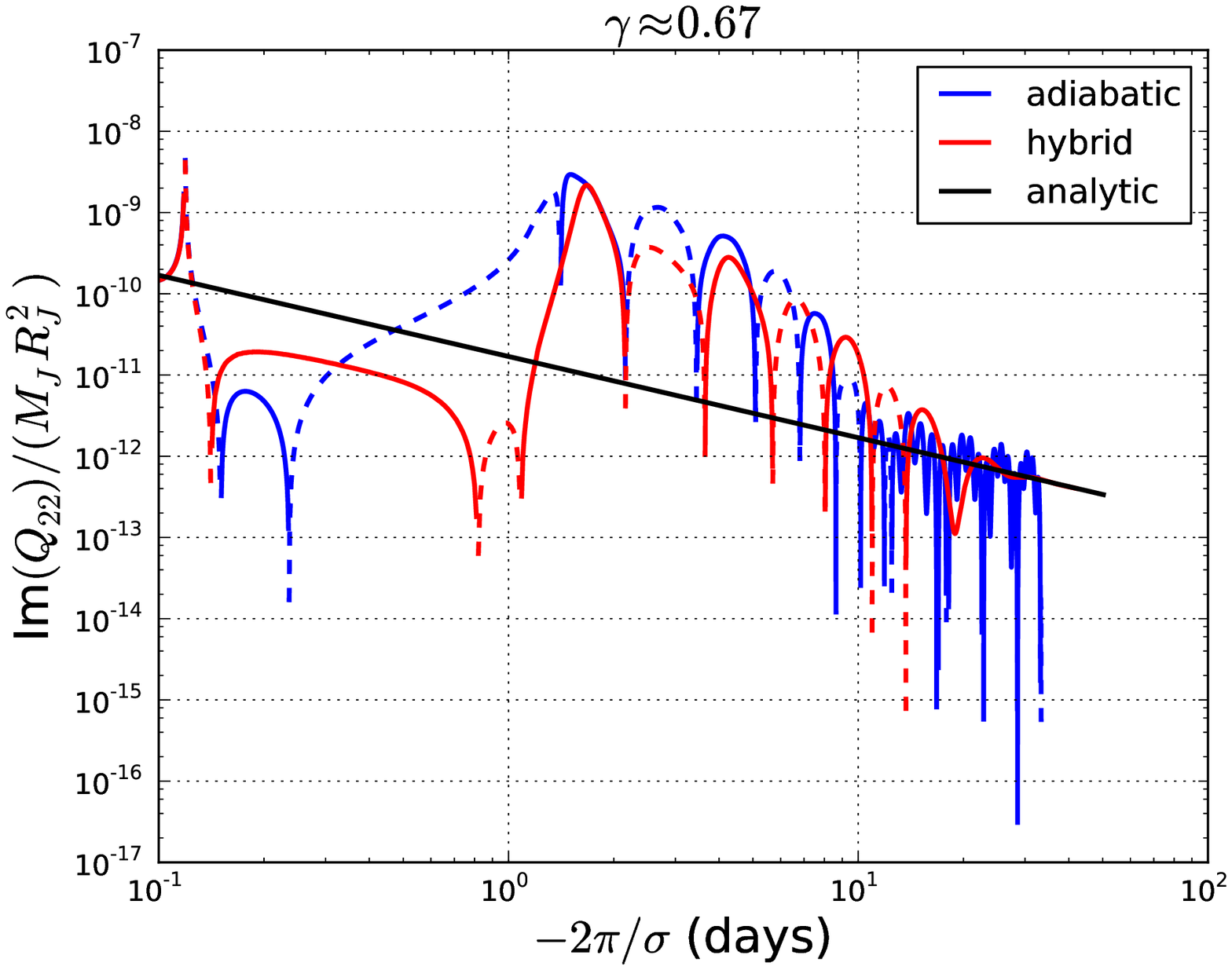}{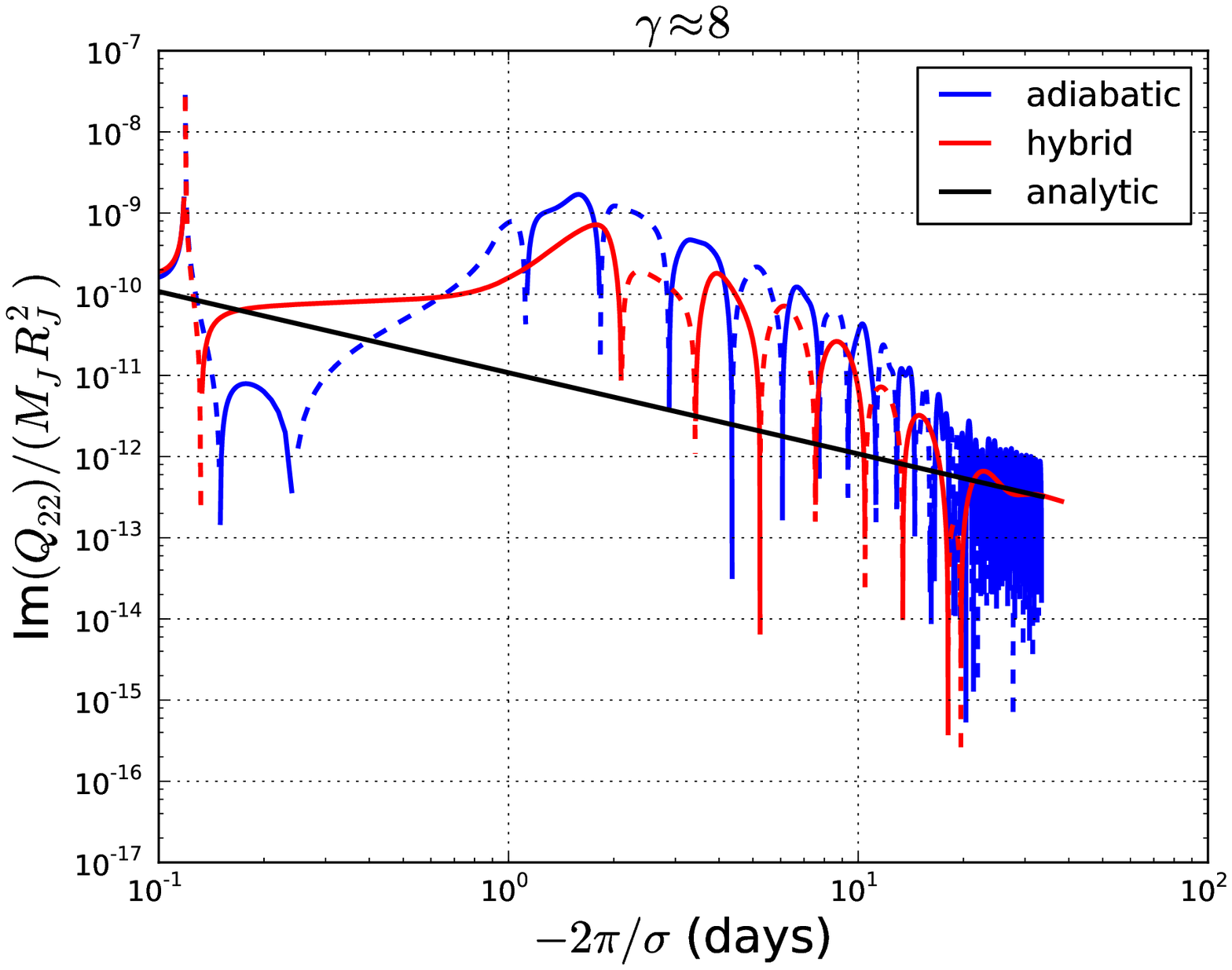}
\caption{The imaginary part of the quadruple moment $Q_{22}$ for the two cases of opacity ratios. Both adiabatic and our hybrid models are presented in each case for comparison. The solid (dashed) line indicates positive (negative) value, implying the planet is away (towards) from the spin-orbit synchronization by the torque acting on the thermal bulge.}
\label{fig2}
\end{figure}


\begin{figure}
\begin{center}
\includegraphics[scale=0.5]{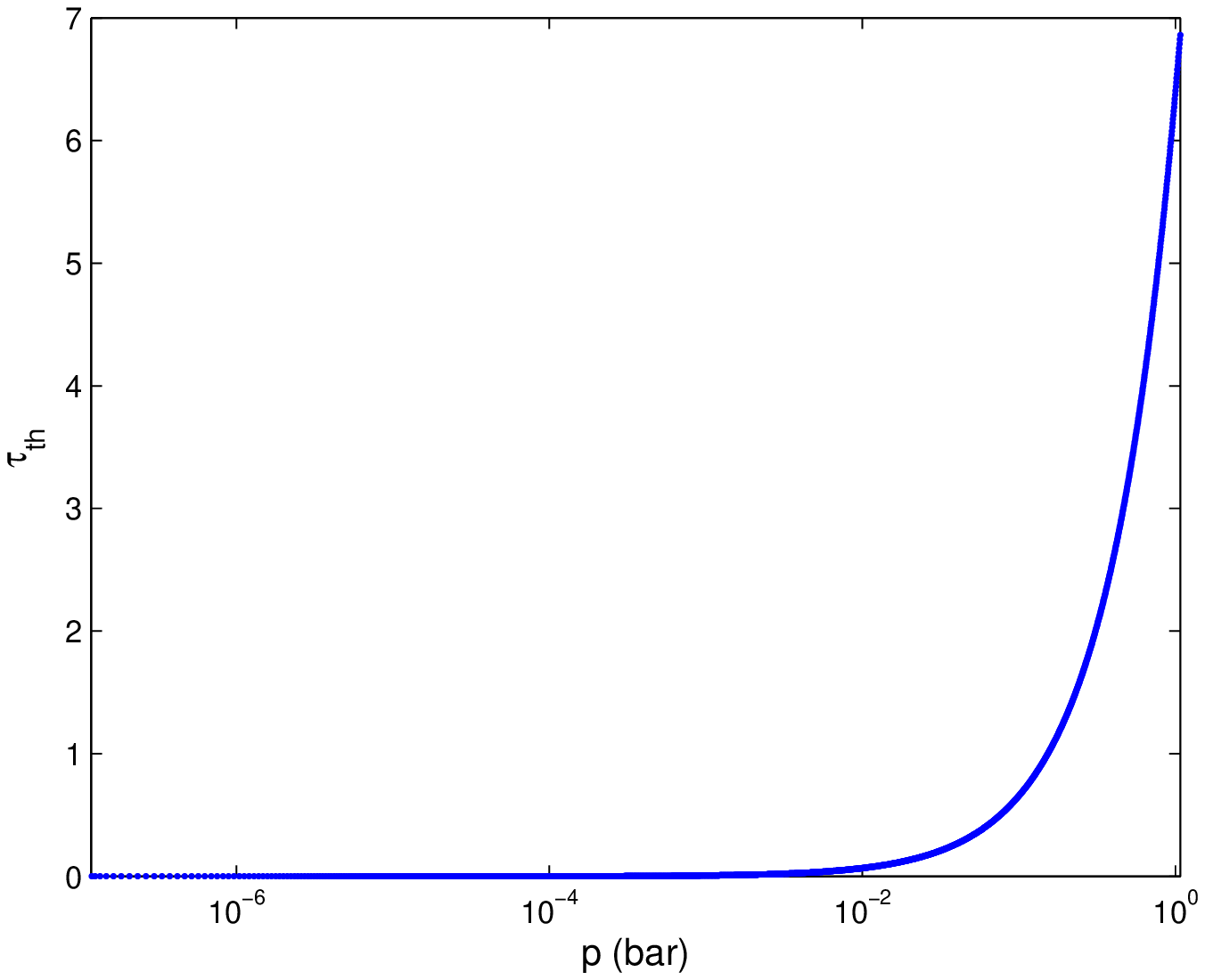}\qquad \includegraphics[scale=0.5]{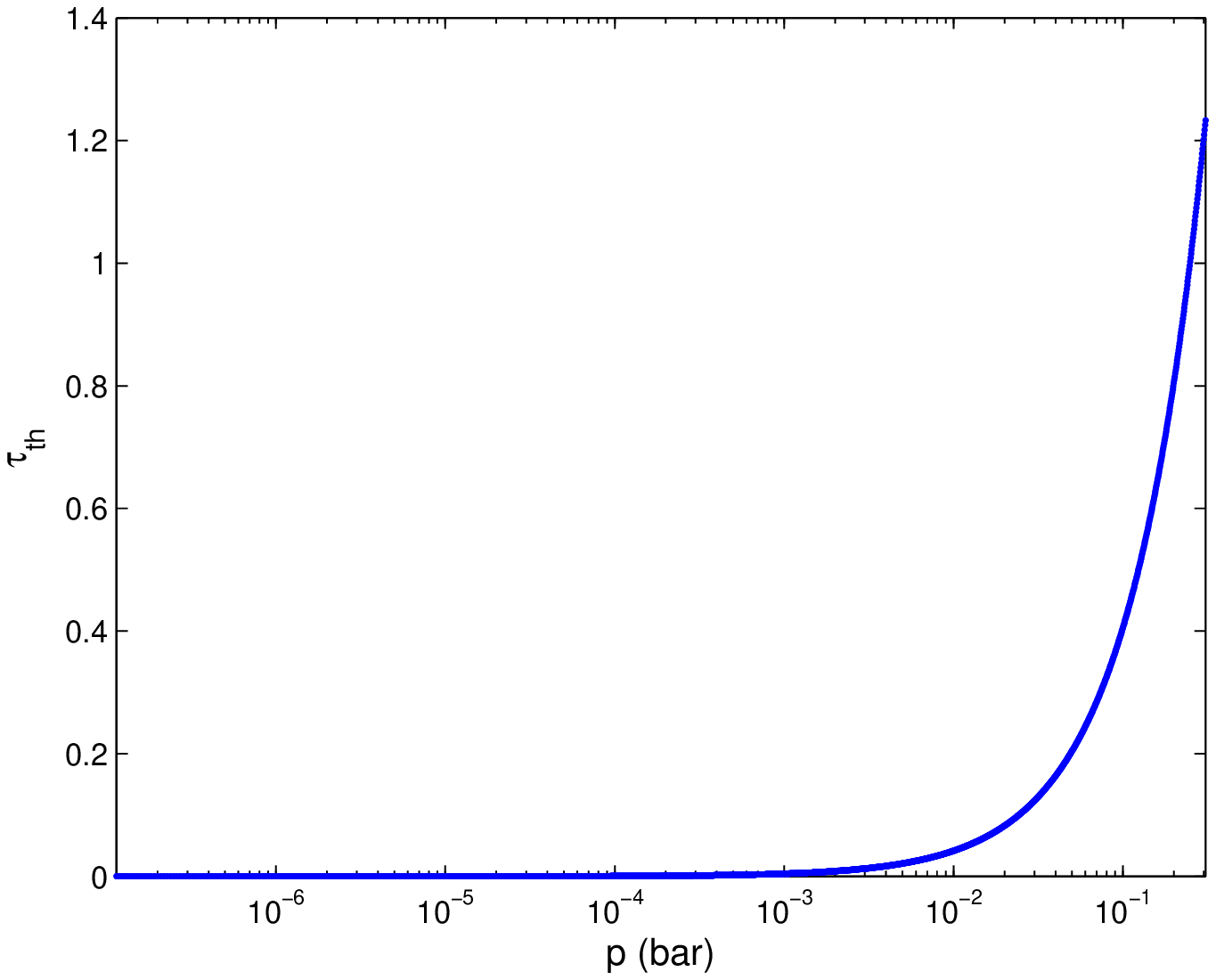}\\
\includegraphics[scale=0.5]{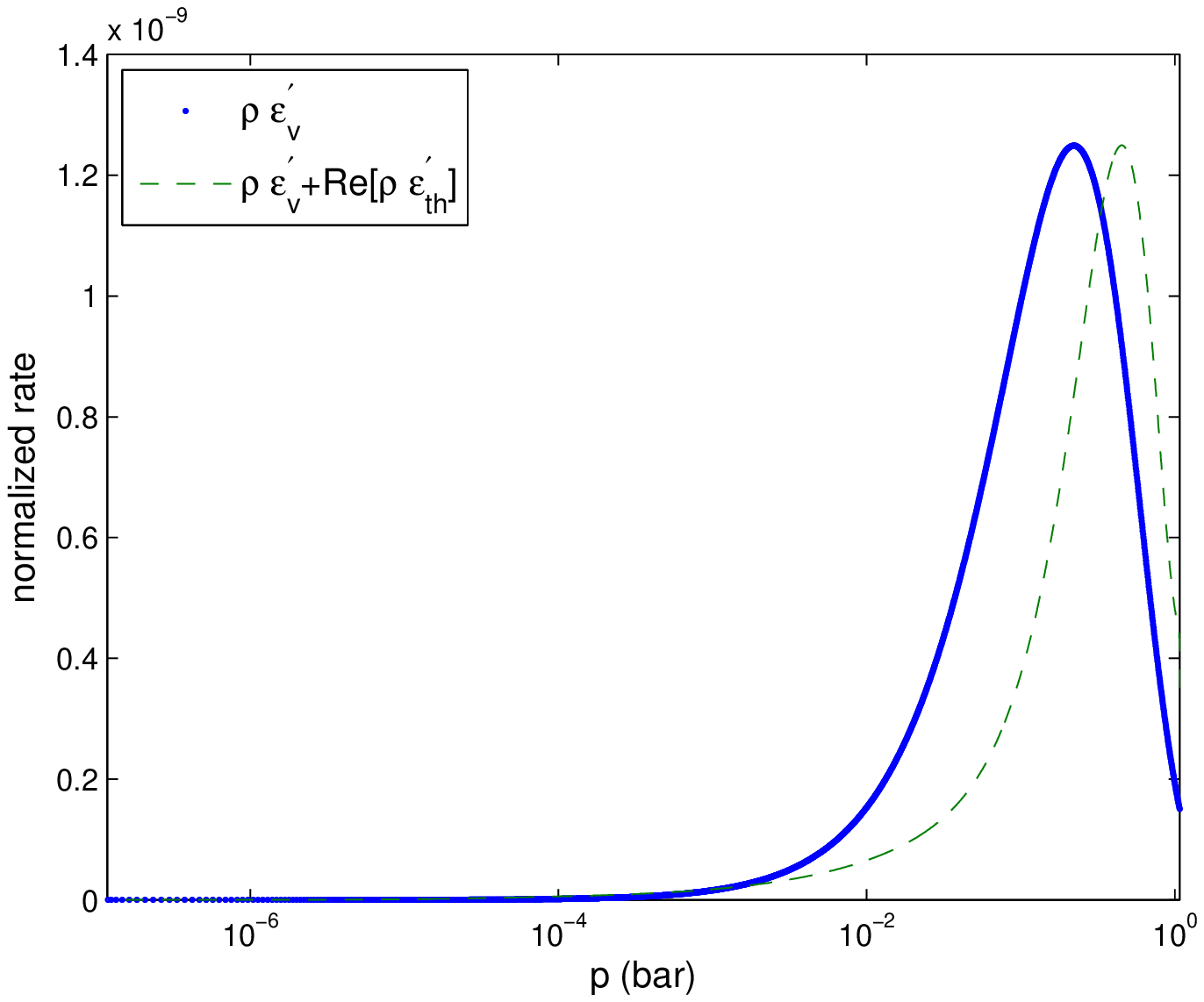}\qquad\includegraphics[scale=0.5]{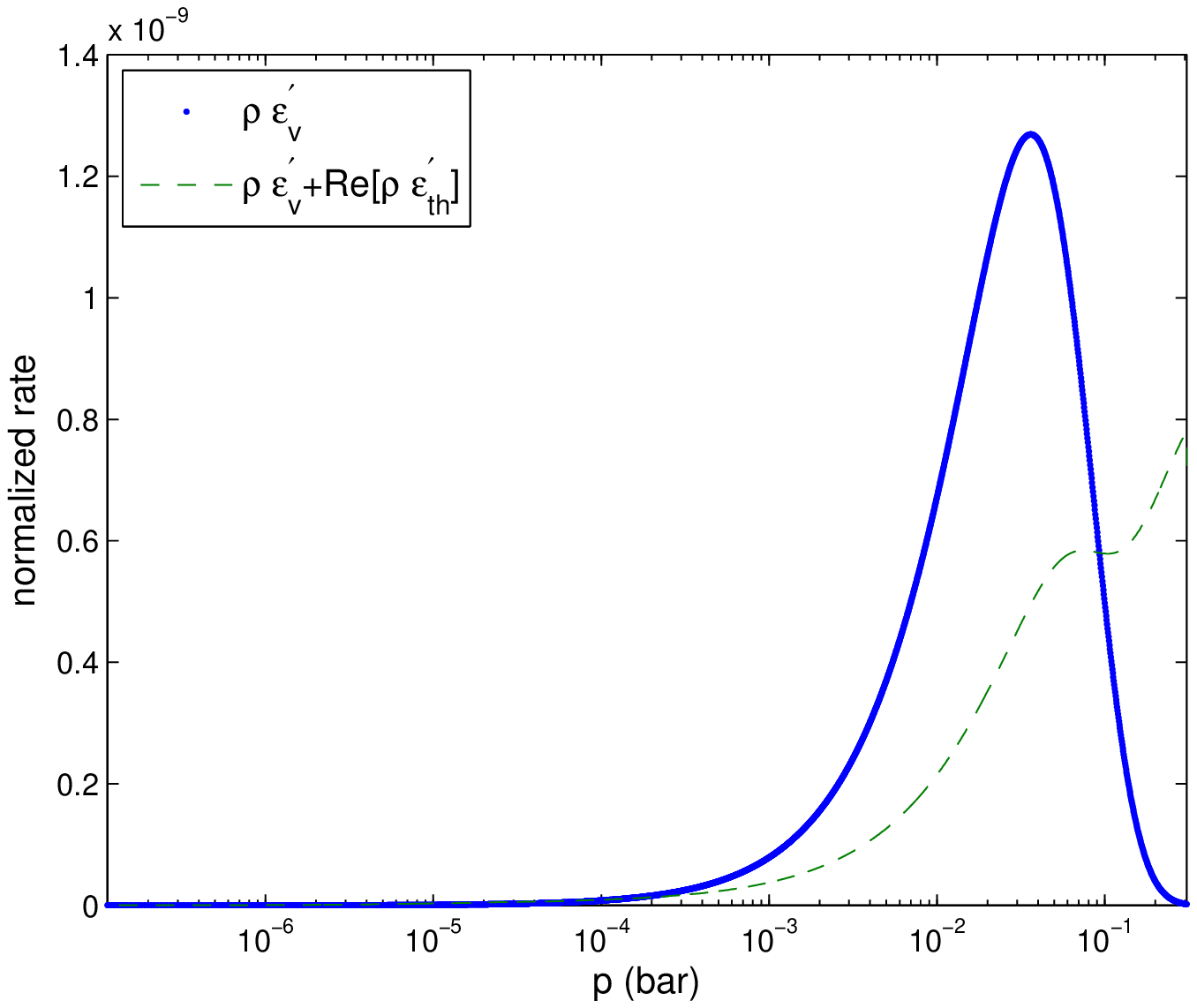}\\
\includegraphics[scale=0.5]{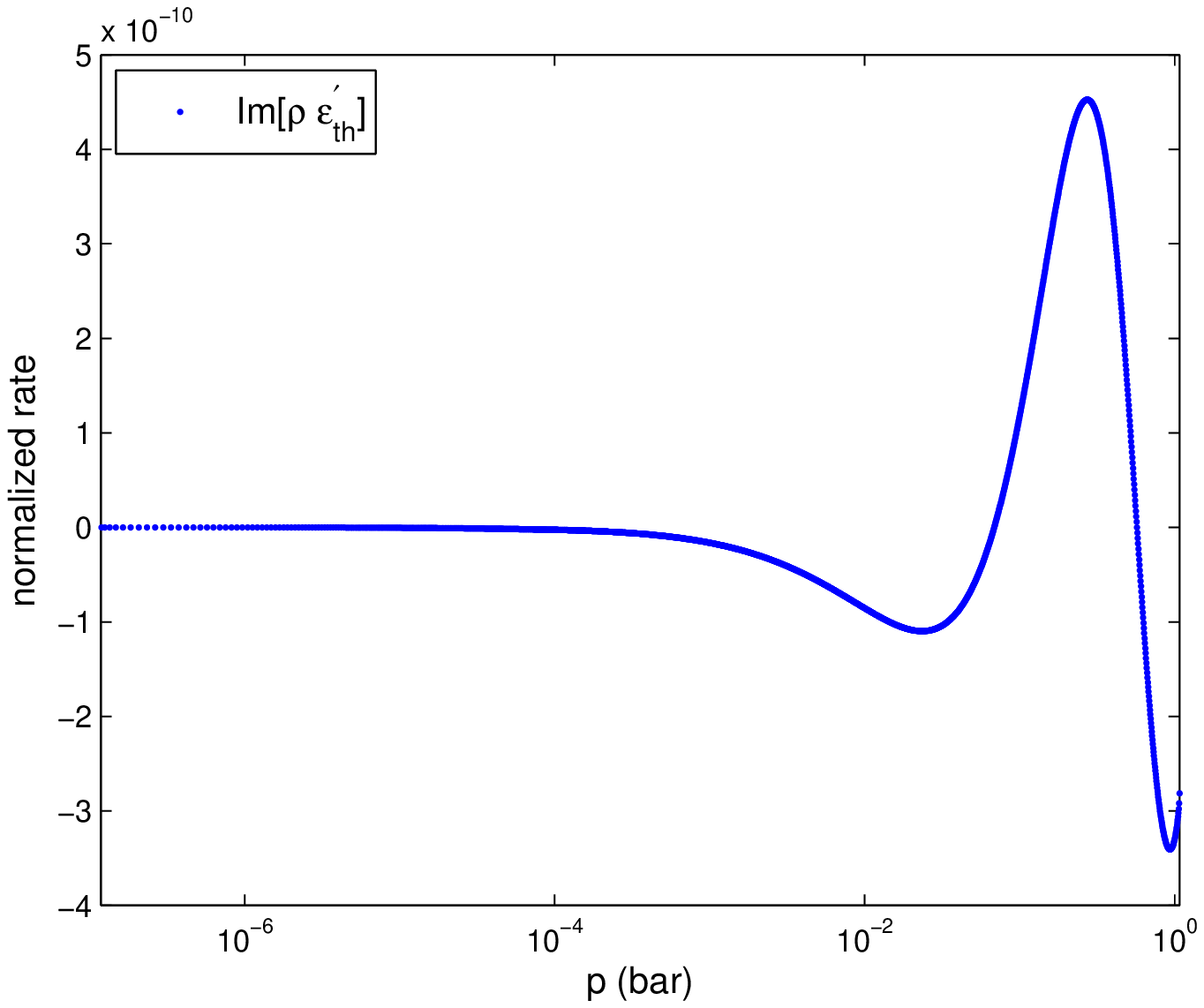}\qquad\includegraphics[scale=0.5]{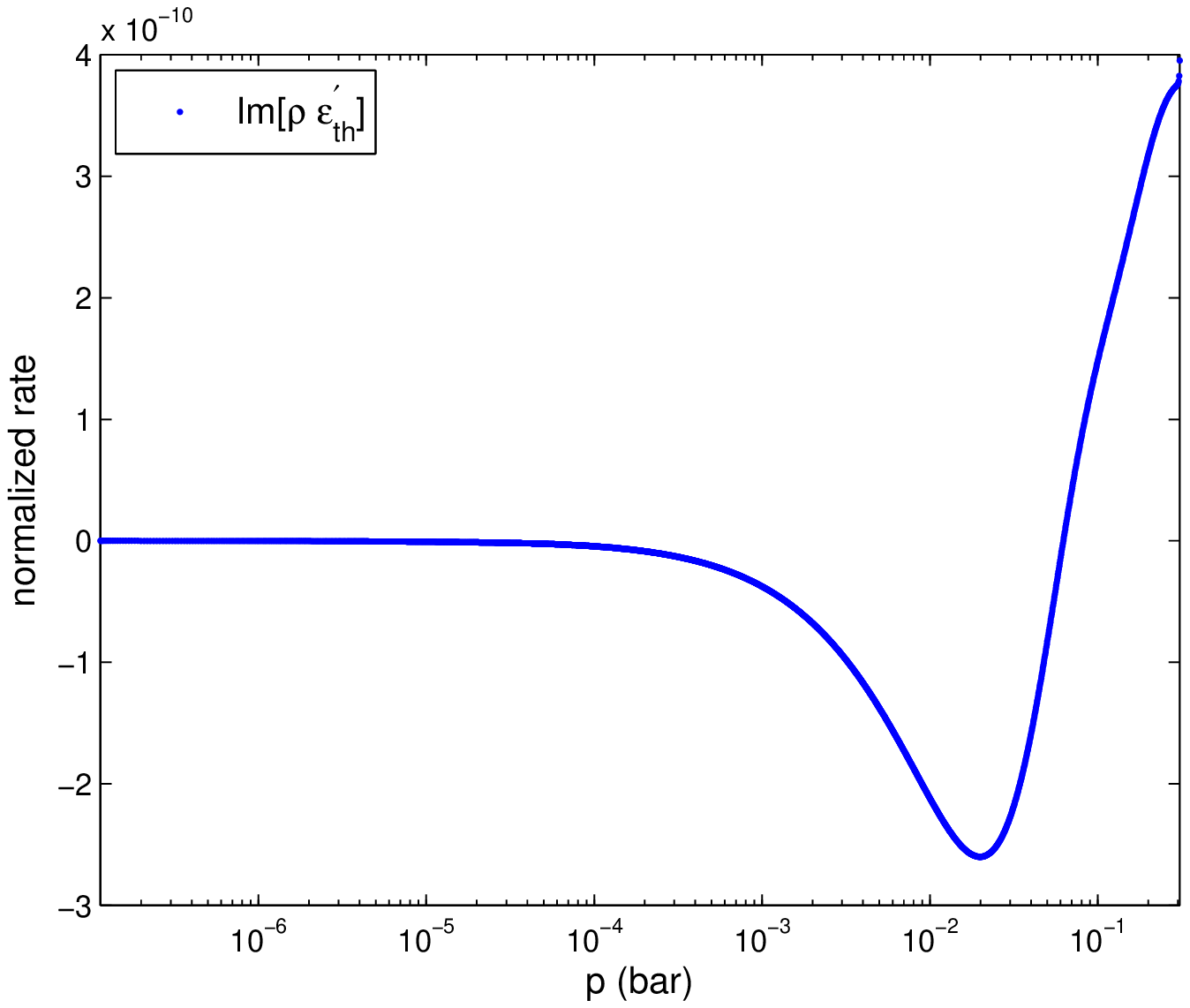}
\end{center}
\caption{Radial profiles of
the thermal optical depth $\tau_{th}$ (top panel), the perturbations of visible radiative heating $\rho \epsilon'_v$ and its value plus the real part of thermal radiative heating/cooling $\rho \epsilon'_v +$Re[$\rho \epsilon'_{th}$] (middle panels), and  the imaginary part Im[$\rho \epsilon'_{th}$] (bottom panels) in the atmosphere when $2\pi/\sigma=-2$ days. Two opacity cases are presented as follows: $\gamma=0.67$ (left panels) and 8 (right panels). There are the same as the cases shown in Figure~\ref{fig2}.}
\label{fig3}
\end{figure}

 In Figure~\ref{fig2}, we compare the Im[$Q_{22}$] spectrum in our hybrid model (in red) with that in the adiabatic model calculated by AS (in blue)
for the two opacity cases. The adiabatic results are similar to the result computed in AS (see their Fig 5) despite different background states being applied to HD 209458 b. To further clarify the fidelity of our calculation,  we multiply $t_{th}$ in the atmosphere and $1/\eta$ in the interior by a large number to artificially suppress the non-adiabatic effects. We subsequently verify that the curve in the hybrid model almost overlaps with that in the adiabatic model (not shown here).  Furthermore, Figure~\ref{fig2}  shows that the curves in the two models merge when the forcing period is close to $-0.1$ day; i.e. the forcing timescale is significantly smaller than the radiative timescale, resulting in adiabatic perturbations. In the other extreme, when the forcing period is large, the curves from two models approach the analytic solution (denoted by the black line, see eq.\ref{eq:analytic}) associated with the quasi-equilibrium thermal tides near the synchronous state \citep{AL18}.

Generally speaking, Figure~\ref{fig2} shows that the spectrum of Im[$Q_{22}$] in the hybrid model is comparable to that in the adiabatic model for the two opacity cases. More precisely, the figure shows that in the case of $\gamma=8$, the spectrum of Im[$Q_{22}$] for g-modes of forcing periods $\gtrsim 1$ day in the hybrid model is overall slightly smaller than that in the adiabatic model. This is suggestive of a weaker greenhouse effect. 
Figure~\ref{fig3} plots the radial profiles of thermal optical depth and the source terms of the energy equation shown in eq(\ref{eq:rad_perb}) when the forcing period is $-2$ days and close to the natural frequency of the lowest g-mode as shown in AS. The source terms are plotted in the same normalized unit of our calculation. The maximum pressure level in the plots lies at the bottom of the atmosphere, approximately corresponding to the base of the stellar heating layer.
As expected, the visible stellar heating peaks at a larger pressure level and thus larger $\tau_{th}$ in the case of $\gamma \approx 0.67$ than in the case of $\gamma \approx 8$.
We observe from the middle panels of the figure that the total heating (in  green), manifested by the sum of the visible and real part of infrared terms (i.e., $\rho \epsilon'_v + {\rm Re}[\rho \epsilon'_{th}]$), has a positive real value (indicative of heating) and falls deeper than the visible stellar heating profile (in blue) towards the optical thicker region in both opacity cases. It is because as the incoming visible radiation decays with depth due to absorption characterized by $\kappa_v$, the infrared radiation is generated with the efficiency characterized by $\kappa_{th}$. Resultantly, the infrared heating in the case of $\gamma \approx 0.67$  is stronger and is located at an optically thicker region than in the case of $\gamma \approx 8$. In other words, a weaker greenhouse effect occurs in the case of $\gamma \approx 8$ and thus explains why the Im[$Q_{22}$] spectrum is generally smaller than that in the adiabatic case as shown in Figure~\ref{fig2} .
 The imaginary parts of the thermal terms are also plotted in the bottom panels for two opacity cases. These terms introduce an additional phase shift for the entropy perturbations (and thus the resulting tidal waves) relative to those in the adiabatic model.
 
\subsection{Contribution to quadrupole moment in relation to planet structure}
 
\begin{figure}
\plotone{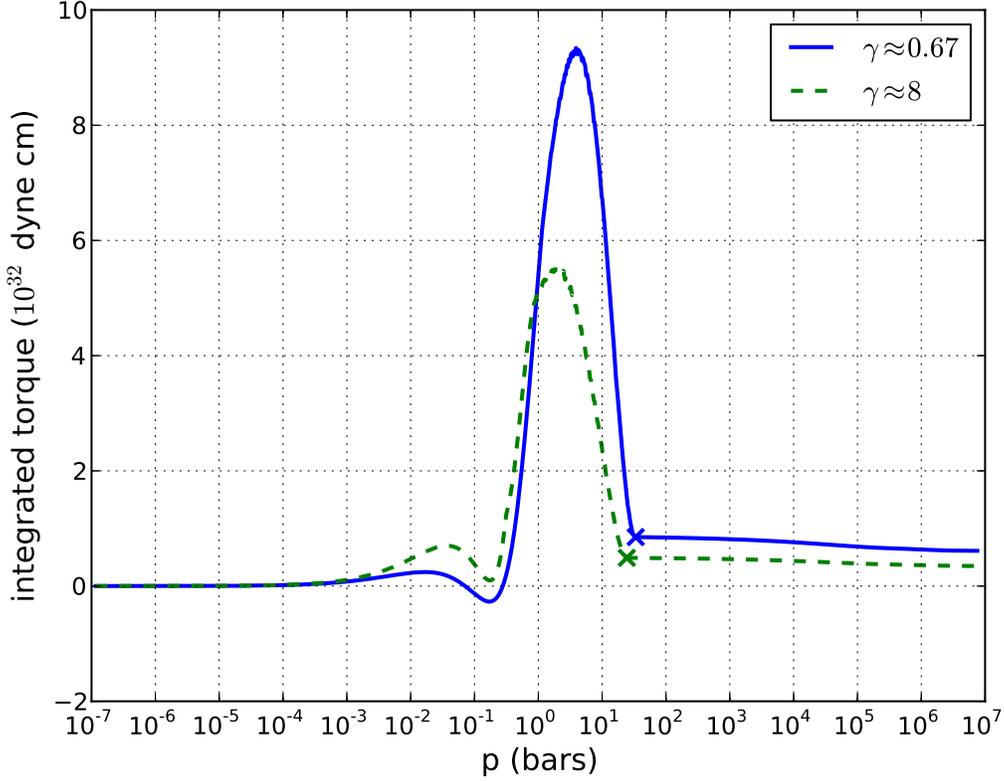}
\caption{The torque integrand due to thermal bulges in eq(\ref{eq:torque}) integrated from outside in for $\gamma \approx 0.67$ (solid line) and 8 (dashed line) in the case of forcing period equal to $-2$ days. The crosses indicate the locations of rcb.}
\label{fig4}
\end{figure} 
 
We also identify the location in the planet that contributes most to the thermal quadrupole moment. In the limit of small $\sigma$, the integrand of eq(\ref{eq:analytic}) suggests a signifiant contribution from the region where the radial profiles of $\sigma^2/N^2$ and the stellar heating $\rho \Delta S$ significantly overlap. For a large $\sigma$, we consider $-2$ days as a representative demonstration again and plot the integrand of eq(\ref{eq:torque}) integrated radially from outside in (i.e., integrated from $r_{top}$ to $r$)  in Figure~\ref{fig4} to reveal the torque contribution. We observe that in both opacity cases, the torque (and by extension the quadrupole moment) is small from high altitudes and then starts to steeply increase beneath the stellar heating layer (i.e., $\gtrsim 1$ bar for $\gamma \approx 0.67$ and $\gtrsim 0.1$ bar for $\gamma \approx 8$). Integrated further deeper into the radiative interior, the torque decreases rapidly with depth until reaching the radiative-convective boundary (rcb) that is marked by the cross in the figure. The peak of the integrated torque in the case of $\gamma \approx 0.67$ lies at a greater depth than that in the case of $\gamma \approx 8$ due to the deeper deposition of stellar heat.

Physically, the waves driven by the stellar heating in the atmosphere excites the quadrupole moment that is largely cancelled by the quadrupole moment contributed by the radiative interior below the atmosphere. The curve of the integrated torque continuously, but comparatively more slowly, decreases with depth below the rcb; this indicates that the evanescent wave in the convective interior continues to reduce the quadrupole moment at a slower rate. 
While the integrated torque in the radiative layer change dramatically, the net torque over the radiative envelope is small. This is indicated by the value at the rcb (the cross), relative to the peak value at $\sim 0.1-1$ bar. Since the net contribution to the quadrupole moment from the g-mode in the atmosphere and radiative zone is small, the contribution from the evanescent wave in the convection zone is not insignificant as illustrated in Figure~\ref{fig4}. In other words, by way of the evanescent wave, the convective zone can couple with the region above the rcb to form a planetwide thermal bulge.

\begin{figure}
\plottwo{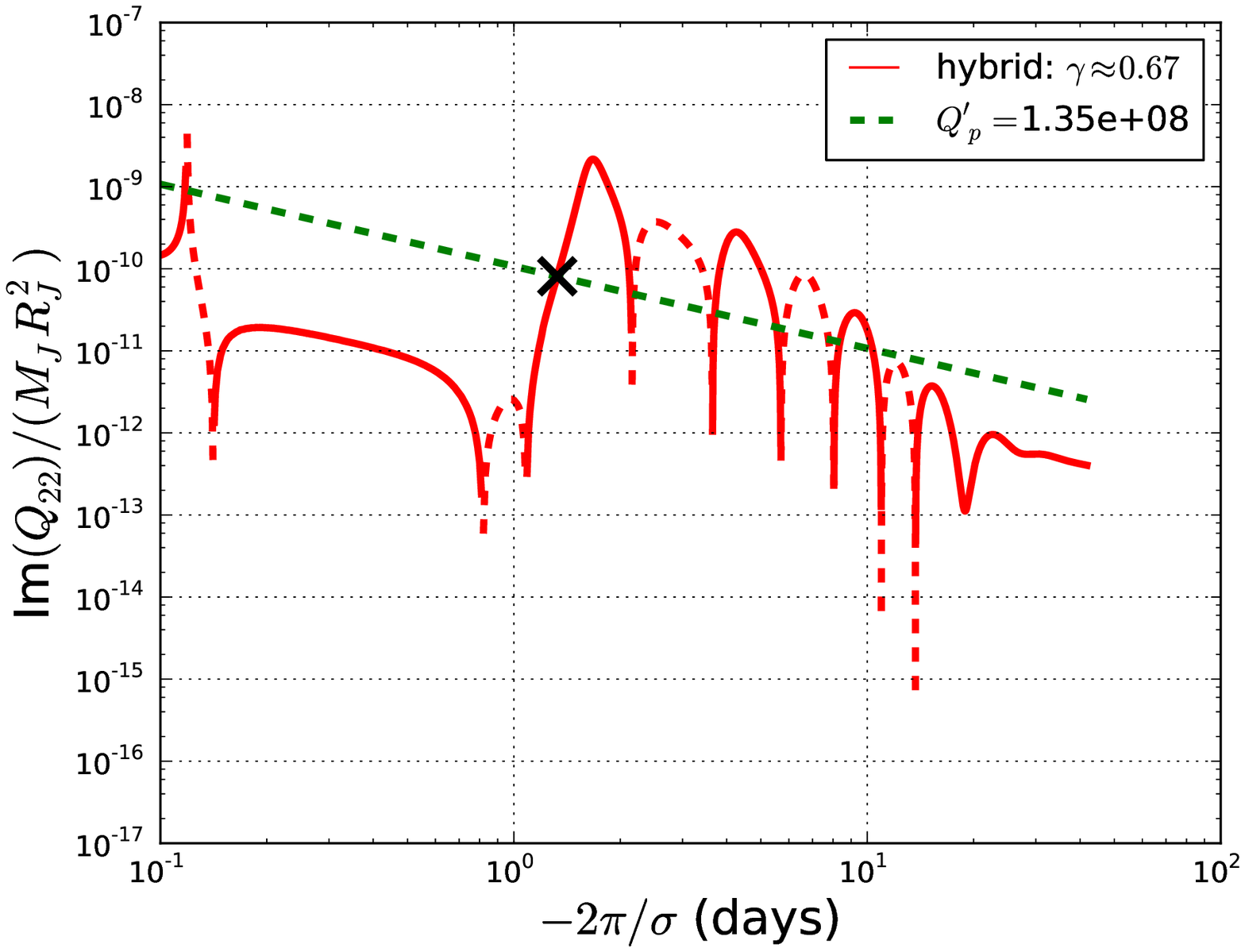}{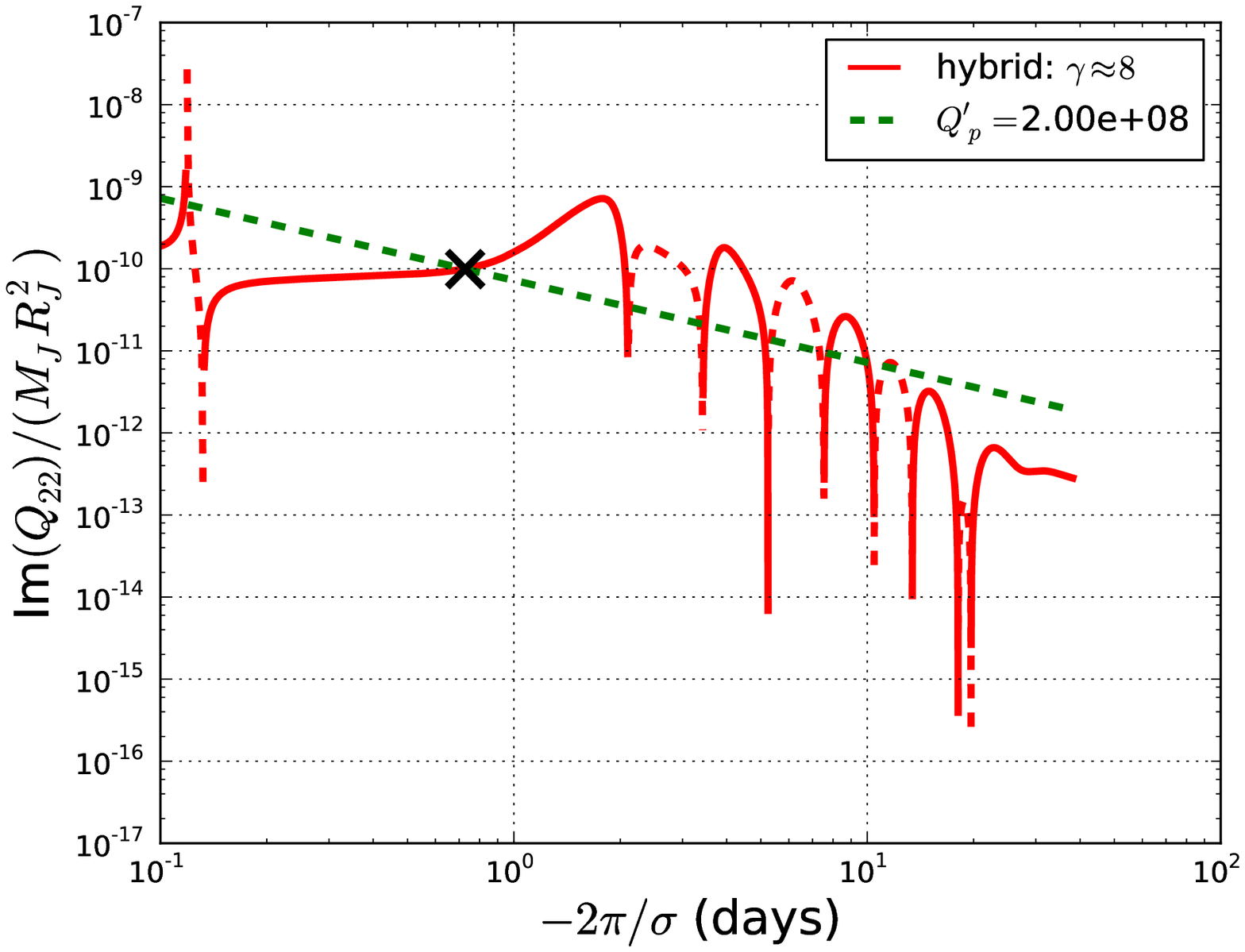}
\caption{The imaginary part of the quadrupole moment excited by thermal dynamical tides (red curve, see Figure~\ref{fig2}) overplotted with that driven by gravitational equilibrium tides (green curve) with the quality factor $Q'_p=1.35\times 10^8$ for $\gamma \approx 0.67$ (left pane) and  $Q'_p=2\times 10^8$ for $\gamma \approx 8$ (right panel). One of the intersections between the two curves, marked by a cross, gives the stable solution to provide the necessary tidal heating for the planet inflation.}
\label{fig5}
\end{figure}

\subsection{Tidal quality factor}

With the solutions for Im[$Q_{22}$] as a function of the forcing period $\sigma$,  we can specify the asynchronous state of the planet using the model of gravitational equilibrium tides. Figure~\ref{fig5} shows the frequency spectrum of Im[$Q_{22}$] contributed from the thermal dynamical tides in our hybrid model (red curve) and that from the gravitational equilibrium tides (green curve, see eq.\ref{eq:T_grav}). The stable torque balance occurs when the green curve intercepts the red solid curve that increases with the forcing period. We adjust the value $Q'_p$ to move the green curve up and down in the plot until the stable dynamical solution at an intercept provides the necessary internal tidal heating to sustain the observed planet radius from the MESA code. Thus, we determine the solution indicated by the black cross in the figure. Namely, $Q'_p \approx 1.35 \times 10^8$ and the forcing period is approximately $-1.3$ days in the case of $\gamma \approx 0.67$ (left panel) and  $Q'_p\approx 2 \times 10^8$ and the forcing period $\approx -0.7$ days in the case of $\gamma \approx 8$ (right panel).
Given the orbital period of 3.52 days, the spin periods are thus approximtely 1.5 and 1 days for $\gamma=$ 0.67 and 8, respectively. Note that the corresponding synchronization timescale by the gravitational tides with $Q'_p\approx 10^8$ can be merely $\sim$ a few million years according to the torque in eq(\ref{eq:torque}) if there is no other mechanism against it, such as a thermal bulge in this study. 

Moreover, it can be observed from Figures~\ref{fig5} and \ref{fig2} that the slope of the Im[$Q_{22}$] for the gravitational equilibrium tide is the same as that for the thermal tides in the limit of small $\sigma$. Both are $\propto \sigma$ to be related to the constant time lag, as has been pointed out by \citet{AL18}.

\subsection{Two-stream approximation vs. Newtonian cooling}
\begin{figure}
\plottwo{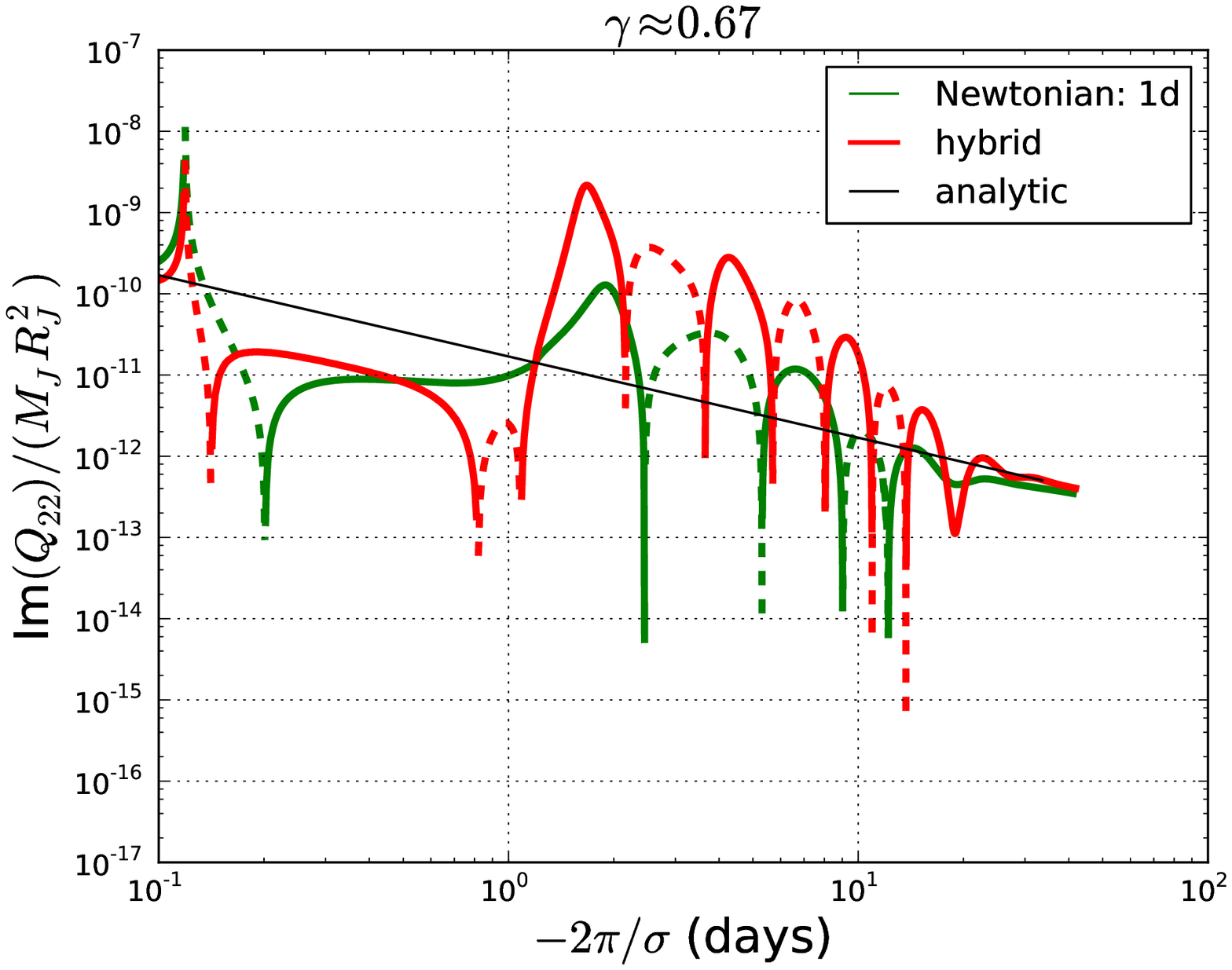}{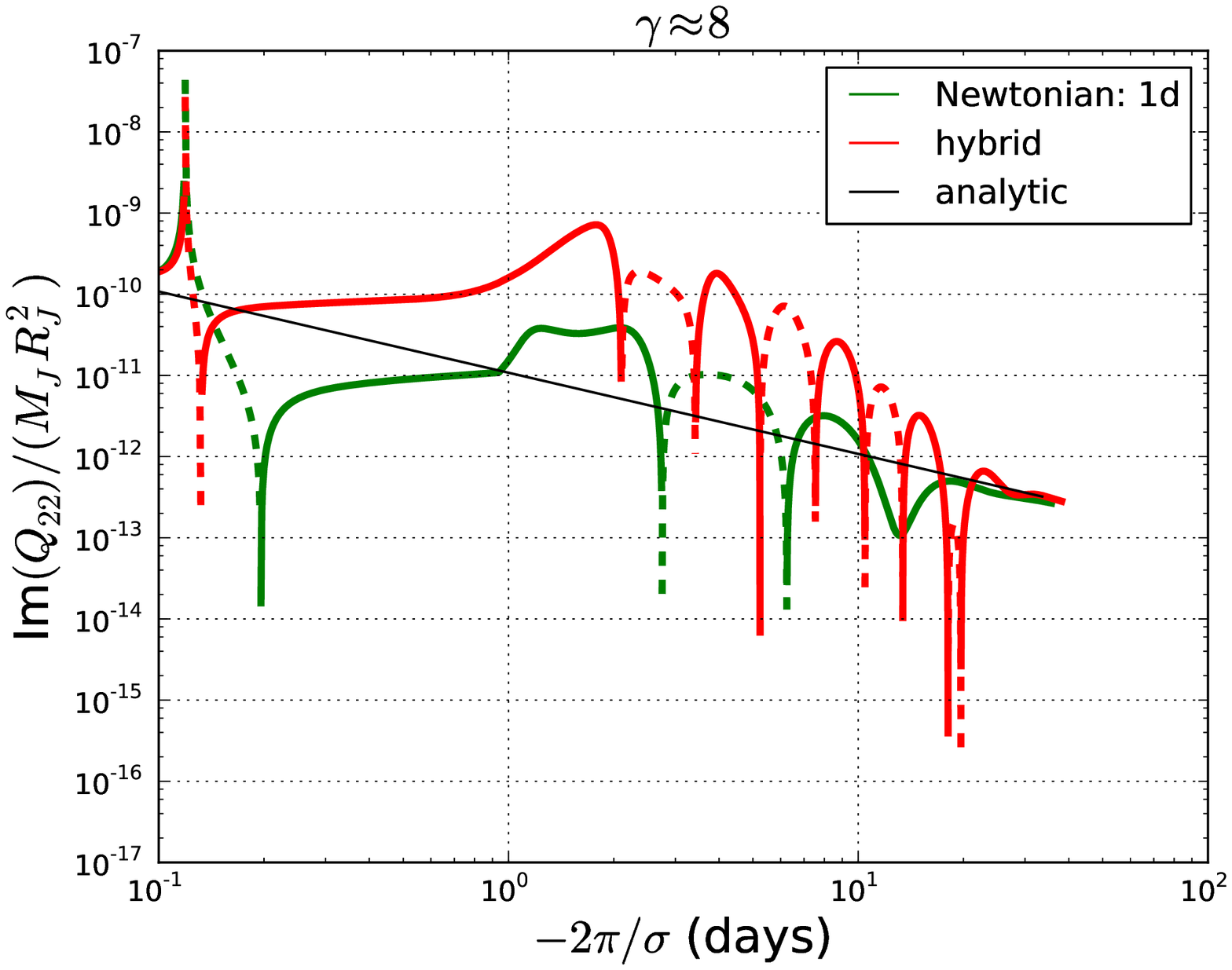}
\caption{The spectra of Im[$Q_{22}$] in the Newtonian damping model by \citet{AL18} with $t'_{th,*}=1$ day and in our hybrid model. As shown in the previous figures, the following two opacity cases are presented: $\gamma \approx 0.67$ (left panel) and 8 (right panel).} 
\label{fig6}
\end{figure}

Based on a prescribed background state and the adiabatic analysis in AS, \citet{AL18} incorporated the effect of Newtonian damping in calculating the gravitational torque on the semi-diurnal component of thermal bulge in the atmosphere of an asynchronously rotating hot Jupiter. In this model, the non-adiabatic term $\nabla \cdot {\bf F'}$ in eq(\ref{eq:perb3}) is simply replaced by $-(1/t_{th})\Gamma_1 p (T'/T)$.
Motivated by the radiative transfer simulations conducted by \citet{Iro}, the authors adopted the pressure-dependent radiative timescale $t'_{th}$  ($=2\pi t_{th}$) of the form $t'_{th,*}/2 [ (p/p_*)^{1/2}+(p/p_*)^2]$ with the parameters $t'_{th,*} \sim 1-10$ days and $p_* \approx 1$ bar at the base of the stellar heating layer in the case of HD 209458 b. In our model with the two stream approximation, $\kappa_{th}$ is a constant parameter throughout the atmosphere overlying an optically thick interior modeled by radiative diffusion. The radiative timescale from eq(\ref{eq:t_th}) based on the two stream formulation is also $\sim 1$ day in the atmosphere and is expected to increase with depth  where radiative diffusion is at work in our model. Thus, we apply their prescription for the Newtonian damping to our background state from the MESA code and calculate the Im[$Q_{22}$] spectra. The purpose is to gather a rough sense of the difference between the two approaches. The results are shown in Figure~\ref{fig6}. The curves in the two models merge in the adiabatic regime when the forcing period is approximately 0.1 days and thus\ $\sigma t_{th}$ is large. The two curves approach the analytic solution in the limit of small $\sigma t_{th}$ beyond the forcing period $\sim 30$ days. For the forcing period in between, the strength of Im[$Q_{22}$] in the Newtonian damping model with $t'_{th,*}=1$ day is overall smaller than that in our hybrid model by one order of magnitude; this suggests the importance of the perturbation due to the absorption of the outgoing thermal emissions. 

\section{Summary and Discussions} \label{sec:highlight}
We extend the thermal-bulge problem studied in AS by including the entropy perturbation of the thermal radiation in the infrared in the framework of the two-stream approximation for the atmosphere, which is connected to the perturbations in the optically thick planet interior to construct a hybrid model for thermal tides. The background state is provided from the MESA code for the interior and eq(49) in \citet{Guillot} for the atmosphere. We consider the atmospheric structure in the two opacity cases; i.e. $\gamma \approx 0.67$ and 8. The former introduces a stronger greenhouse effect, whereas the latter exhibits a temperature inversion. A linear theory is conducted and applied to the hot Jupiter HD 209458 b for a fiducial study. 

\subsection{Less damped thermal bulge due to greenhouse}

We observe that our result does not differ significantly from the result from  AS despite the radiative timescale ($t'_{th} \lesssim 1$ day) being smaller than the dynamical timescale associated with lower-order g modes (a few days). It is because the perturbation of absorption of the outgoing thermal radiation contributes the heating against the thermal escaping emissions (i.e. Newtonian damping).
In the case of $\gamma \approx 0.67$, the total heating including the infrared part of heating and damping penetrates slightly deeper than the visible part of the stellar heating alone, thus exhibiting a stronger greenhouse effect than in the case of $\gamma \approx 8$. We also find that for the thermal tides excited by the g-mode of lowest order, the contribution to the quadrupole moment from the convection zone is not insignificant since the net contribution from the atmosphere and radiative zone is small. Evidently, the thermal bulge extends throughout the entire planet in our case studies.

We also compare the results with the recent results by \citet{AL18} who extensively studied the influence of Newtonian damping.
Given the radiative timescale about 1 day at the level of the heating layer,  the quadrupole moment in our hybrid model is larger than that in their model by one order of magnitude, implying the suppression of Newtonian damping by the greenhouse effect. However, we should be cautious about the comparison. The atmospheric $\kappa_{th}$ is constant in the two stream approximation and thus $t'_{th}$ from eq(\ref{eq:t_th}) varies with  $T^{-3}$ in the atmosphere, which does not vary as steeply as the pressure dependence $\propto p^{1/2}$ above the level $p_*=1$ bar adopted by \citet{AL18}.  This could imply that in the case of  $\gamma \approx 8$, the waves excited by the stellar irradiation at higher altitudes are not significantly damped by Newtonian damping in our model. 

In any case, we present a theoretical framework of thermal tides to bridge the two observational properties, i.e. hot-Jupiter inflation and thermal inversions. The study of thermal inversions of atmospheres has been an active field to better understand thermal  properties and chemical abundance, such as the C/O ratio, thus also relevant to the formation and evolution of hot Jupiters \citep[e.g.,][]{Madhus,GM19}.
As a wealth of studies of thermal inversions on hot Jupiters have been underway, the atmospheric opacities of these highly irradiated planets are expected to be soon improved.
Future investigations on the two-stream approximation with altitude-dependent thermal and even visible opacities would be informative for the study \citep[e.g.,][]{Lee18}.

\subsection{Uncertainties in modeling the hybrid transition and Eddington coefficients}

In our hybrid model, the connection of linear equations at the interface between the atmosphere and interior is performed to ensure the same values of the perturbations $y_p$, $\xi_r$, $F'_r$, and $y_T$ at the interface from both sides. The procedure is probably oversimplified but enables us to make a first attempt at deriving the linear equations to be more consistent with a more realistic background state from MESA. A more careful treatment for the connection algorithm such as a smooth transition across the interface should be considered to refine the results.

To deal with the absorption of reprocessed stellar irradiation in the near-infrared, we attempted to perturb the radiative transfer equations in addition to the dynamical equations. We apply
the same Eddington's coefficients among the moments of the specific intensity of a light ray to their perturbations to close the linear equations without justification. This procedure amounts to the same direction of perturbed radiation fields as that of the background radiation. The assumption will be investigated in a future work. Furthermore, the effect of radiative scattering that is ignored in our model should be taken into account in the Eddington closure scheme in a line of analysis congruent to the one adopted by \citet{Heng}.

\subsection{Outlook of our model}

Under the assumption of the torque balance and thermal equilibrium,
we evaluate the tidal quality factor of the gravitational equilibrium tides to be $\approx 10^8$ for HD 209458 b. The spin period is approximately $1-2$ days, which is a significant departure from the orbital period of 3.52 days but consistent with the frequency associated with the g mode of the lowest order as shown in AS. The $Q'_p$ value could likely be smaller if the planet is more rich in metals \citep[e.g.,][]{Thorn16} or if the tidal heating is deposited at shallower depths \citep[e.g.,][]{Gu04,KY17}. Since our study focuses on the non-adiabatic effects in one hot-Jupiter case, we are not able to justify the general scaling law derived by \citet{Socrates} for the heating efficiency of hot Jupiters in relation to the incident stellar flux, planet radius, and orbital period. 
The scaling law has been convenient to study the scenario of hot-Jupiter re-inflation around evolved stars \citep{LF16} as well as to provide a result to be constrained with the statistical understanding of heating efficiency \citep{TF18}.

However, it should be noted that the scaling law was derived based on the Gold-Soter formulation for equilibrium thermal tides and the background state specific to HD 209458 b, with the purpose of prompt and easy applications \citep{GS69,AS10,Socrates}. 
It remains unanswered whether the scaling law derived from the equilibrium tides for one particular planet can still be an effective approximation for the dynamical tides in general for all hot Jupiters.
For instance, in our linear theory for dynamical thermal tides, everything else being the same, the quadrupole moment $Q_{22} \propto F_* \propto T^4_{eq}$ due solely to thermal excitation. On the other hand in terms of radiative damping, as has been discussed above, the thermal timescale $t_{th} \propto T^{-3}$. Assuming that this timescale can be approximated to be $\propto T_{eq}^{-3}$ and that the greenhouse effect is weak, $Q_{22}$ can be reduced by Newtonian damping as $T_{eq}$ increases.\footnote{On the contrary, the quadrupole moment excited by thermal tides in the Gold-Soter formulation does not increase but decreases with increasing $t_{th}$ \citep[AS,][]{Socrates}, which eventually leads to Socrates' scaling law for dissipation $\propto t_{th} \propto T_{eq}^3$.}  This process has been invoked to explain the observed trend of increasing day-night temperature contrast in hot Jupiters with increasing $T_{eq}$ via increasing radiative damping of thermal tides \citep{Komacek}.

Moreover,
it is conceivable that there exist various states of electron degeneracy in the interior of hot Jupiters with radius ranging from 1 to 2 $R_J$ to inflate differently in response to a given gravitational tidal dissipation.
Last but not least, the gravitational tides are not modeled in this work but are parameterized by the quality factor $Q'_p$. The dynamical gravitational tides would be more significant in generating the torque and dissipation power in a hot Jupiter in the presence of the Coriolis force and a solid core \citep{OL}. The Coriolis effect on the thermal bulge in the atmosphere has been studied by \citet{AL18} under the traditional approximation. Applying a more realistic model with the MESA code such as the approach that we have presented in this work to other hot Jupiters would be beneficial in improving the understanding. However, the present form of the two stream approximation that ignores visible outgoing radiation is less valid for application to ultra-hot Jupiters.

\subsection{Issues of wave feedback to the background}

\begin{figure}
\plotone{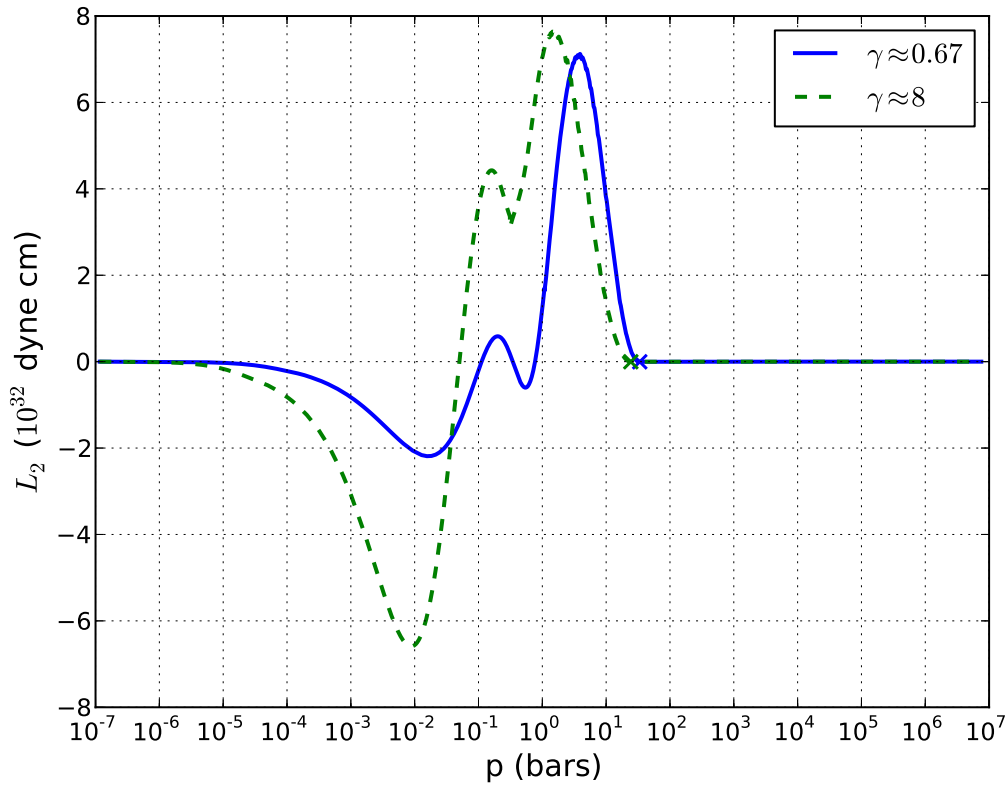}
\caption{The vertical angular momentum flux carried by thermal tidal waves for the forcing period $=-2$ days. The locations of rcb is indicated by the crosses. Two opacity cases are presented.}
\label{fig7}
\end{figure}

Thermal tides have been known as a source of global circulations through wave-mean flow interactions in a hot-Jupiter atmosphere \citep[e.g.,][]{GO,SP,Tsai,Cho,Mayne}. The heat advected by the equatorial super-rotating flow away from the substellar point in some of these wave models is consistent with the infrared phase curve probed towards hot Jupiters \citep[e.g.,][]{Knutson,Zellem}. 
In this work,
we ignored the resulting internal torque and heat due to radiative damping of thermal tides as well as heat advection by global circulations for the sake of simplicity. In the presence of radiative heating and cooling, thermal tidal waves and circulation flows should transport angular momentum and energy between different layers of the planet\citep[][]{Mayne}\footnote{AS considered dissipationless thermal tides. Therefore, there is no internal torque and heat via the dissipation of thermal tidal waves but there exists an external torque on the planet from the central star via the thermal bulge.}. The effects associated with circulations and even inhomogeneous cloud cover \citep[e.g.,][]{SH15,Lines19} are expected to shift the phase and change the amplitude of the thermal bulge that we model in this work. It should be kept in mind that circulations are nonlinear phenomena and therefore are not captured in our linear analysis. 
In addition, most of the global circulation models have been studied for tidally locked hot Jupiters and therefore the outcome of circulations on an asynchronous hot Jupiter remains unclear.

Given all of the complications, we restrict ourselves to approximately evaluate the impact of the feedback due to linear waves for a brief discussion. Using our linear solutions, we can estimate the radial angular momentum flux carried by the semi-diurnal wave through a spherical surface at $r$ \citep{OL,GO}:
\begin{equation}
L_2(r)={2\pi \over \sigma} {\rm Re}[p' (-i \sigma \xi_r)] r^2,
\end{equation}
which is plotted in Figure~\ref{fig7} against gas pressure for the forcing period $=-2$ days. Generally speaking in both opacity cases, the prominent feature of the figure is that the regions associated with the positive gradient of $L_2(r)$ above $\sim 0.01$ bar and below $\sim 1$ bar mostly lose the angular momentum to the stellar heating layer around $\sim 0.1$ bar where the gradient of $L_2$ is almost negative. Thus, in the case of a retrograde thermal forcing ($\sigma<0$), the waves thermally excited from the stellar heating region propagate downward and upward, carrying a positive and negative angular momentum flux, respectively. Resultantly, the layers below and above the stellar heating layer would spin slower and yield radial shear.  The result is reminiscent of the angular momentum transport obtained by \citet{GO}; in their mode,  the retrograde stellar heating imposed at the top boundary excites waves propagating downward and transport angular momentum upwards.
Compared to the case of $\gamma \approx 8$, the peak and trough of the $L_2$ profile in the case of $\gamma \approx 0.67$ shifts slightly downward since the stellar heating penetrates deeper. 
In both opacity cases, the angular moment flux and its gradient below the rcb that is marked by the cross is relatively small as a result of evanescent waves. The magnitude of $L_2$ is on the order of $10^{32-33}$ dyne cm, which is significantly larger than the external torque $T_{grav}$ in eq(\ref{eq:T_grav}) for $Q'_p \sim 10^8$ based on the spin balance from our computed thermal bulge. The corresponding wave energy is in the order of $L_2 \sigma \sim$ a few times 1e26 erg/s. Although this amount of wave energy is comparable to the internal gravitational tidal heating, it is mostly dissipated above the rcb rather than in the convective interior as shown in Figure~\ref{fig7}.
The feedback of the angular momentum and energy by the thermal waves to the background state of the radiative envelope and the atmosphere is substantial. The feedback should be considered in future modeling.

\acknowledgments
We thank P. Auclair-Desrotour and Shang-Ming Tsai for informative discussions at the conference of Extreme Solar Systems IV. We also thank the anonymous referee for useful comments to improve the paper structure. Some of the numerical works are conducted on the high-performance computing facility at the Institute of Astronomy and Astrophysics in Academia Sinica (ASIAA). This research has been supported by the Ministry of
Science and Technology of Taiwan through the grant MOST 105-2119-M-001-043-MY3.

%






\appendix

\section{Linear equations for the atmosphere}
Following \citet{Guillot}, we adopt $f_{Kth}=1/3$ and $f_{Hth}=1/2$ for the Eddington coefficients. 
To solve the linear equations for the atmosphere with a sparse matrix composed of the coefficients of linearized ODEs, eqns(\ref{eq:perb1}), (\ref{eq:perb2}), (\ref{eq:perb3}), (\ref{eq:rad_perb}),   (\ref{eq:RT1}), and  (\ref{eq:RT2}) are rearranged as follows:
\begin{eqnarray}
&&{dy_{p,R} \over dr}+y_{p,R} [f1(r)+F1(r)]+\xi_{r,R} [g1(r)+G1(r)] \nonumber \label{eq:D_first}\\
&& \qquad+ y_{p,I} \mathcal{F}1(r)+\xi_{r,I} \mathcal{G}1(r) + J'_{th,R}\mathcal{H}1(r)+J'_{th, I}\mathcal{J}1(r)=\mathcal{S}1_{R}(r), \label{eq:ODE1}\\
&&{d \xi_{r,R} \over dr}+y_{p,R} [f2(r)+F2(r)] + \xi_{r,R} [g2(r)+G2(r)]  \nonumber \\
&&\qquad +y_{p,I} \mathcal{F}2(r) + \xi_{r,I} \mathcal{G}2(r)+ J'_{th,R}\mathcal{H}2(r)+J'_{th, I}\mathcal{J}2(r)=\mathcal{S}2_R(r),\\
&&{dy_{p,I} \over dr}-y_{p,R}\mathcal{F}1(r) -\xi_{r,R} \mathcal{G}1(r) \\
&&\qquad + y_{p,I} [f1(r)+F1(r)]+\xi_{r,I} [g1(r)+G1(r)] + J'_{th,R} \mathcal{J}1(r) + J'_{th,I} \mathcal{H}1(r)=\mathcal{S}1_I(r), \nonumber \\
&&{d \xi_{r,I} \over dr}-y_{p,R} \mathcal{F}2(r) - \xi_{r,R} \mathcal{G}2(r) \label{eq:ODE2}\\
&& \qquad +y_{p,I} [f2(r)+F2(r)] + \xi_{r,I} [g2(r)+G2(r)]  +J'_{th,R} \mathcal{J}2(r) +J'_{th,I}\mathcal{H}2(r)=S2_I(r), \nonumber \\
%
&&{d H'_{th,R} \over dr} + \alpha_1(r) y_{p,R} + \alpha_2(r) \xi_{r,R} + \alpha_3(r) y_{p,I} + \alpha_4(r) \xi_{r,I} \nonumber \\
&& \qquad + \alpha_5(r) J'_{th,R} + \alpha_6(r) J'_{th,I}= \mathcal{S}3_R(r),\\ 
%
&&{d H'_{th,I} \over dr} + \alpha_7(r) y_{p,R} + \alpha_8(r) \xi_{r,R} + \alpha_9(r) y_{p,I} + \alpha_{10}(r) \xi_{r,I} \nonumber \\
&& \qquad + \alpha_{11}(r) J'_{th,R} + \alpha_{12}(r) J'_{th,I}= \mathcal{S}3_I(r),\\ 
&&{d J'_{th,R} \over dr} = -{\rho \kappa_{th}  \over f_{Kth} }H'_{th,R},\\
&&{d J'_{th,I} \over dr} = -{\rho \kappa_{th}  \over f_{Kth}} H'_{th,I}, \label{eq:D_last}
\end{eqnarray}
where
\begin{eqnarray}
&& t_{th}\equiv \left( {16 \rho \kappa_{th} \sigma_B T^4 \over p} {\Gamma_3 -1 \over \Gamma_1} \right)^{-1},\\
&&f1(r)={\rho g \over p} {1-\Gamma_1 \over \Gamma_1},\\
&&g1(r)={N^2 \rho \over p} - {\sigma^2 \over (p/\rho)},\\
&&f2(r)= {1\over \Gamma_1}-{\lambda\over \sigma^2 r^2}{p \over \rho} ,\\
&&g2(r) = {2\over r}  + {N^2 \over g} - {d\rho \over dp} g,\\
&&F1(r)=
{\rho g \over p}{\Gamma_1 -1\over \Gamma_1}
\left(  {1\over t_{th} \sigma}  \right)^2 {1\over 1+ 1/(t_{th}\sigma)^2} ,\\
&&G1(r)=
-{\rho N^2 \over p}
\left(  {1\over t_{th} \sigma}  \right)^2 {1\over 1+ 1/(t_{th}\sigma)^2},\\
&&\mathcal{F}1(r)=-{\rho g \over p}{\Gamma_1 -1\over \Gamma_1}{1\over t_{th}\sigma}
{1\over 1+ 1/(t_{th}\sigma)^2} ,\\
&&\mathcal{G}1(r)={\rho N^2 \over p}
{1\over t_{th}\sigma}
{1\over 1+ 1/(t_{th} \sigma)^2} ,\\
&&\mathcal{H}1(r)=-{1\over \sigma_B T^4} {\pi \rho g \over 4p}
 {1 \over( t_{th} \sigma)^2}
{1\over 1+ 1/(t_{th} \sigma)^2}  \\
&&\mathcal{J}1(r)= {1\over \sigma_B T^4}  \left[ {\pi \rho g \over 4p}{1 \over t_{th} \sigma} 
-{\pi \rho g \over p} {1\over t_{th} \sigma}
{1\over 1+ 1/(t_{th} \sigma)^2} {1\over 4} \left( {1\over t_{th} \sigma} \right)^2 \right], \\ 
&&\mathcal{S}1_{R}(r)={\rho g \over p}{1\over t_{th} \sigma}
{1\over 1+ 1/(t_{th} \sigma)^2}  \left[ {\rho \over p \sigma} {\Gamma_3 -1 \over \Gamma_1} c_{m=2,n} \kappa_*  F_* \exp(-p\kappa_*/g) \right], \label{eq:S1_R}\\
&&\mathcal{S}1_{I}(r)={\rho^2 g \over p^2 \sigma} {\Gamma_3 -1 \over \Gamma_1}c_{m=2,n}\kappa_* F_* \exp \left(- {p \kappa_* \over g} \right) \left[ 1-{1\over 1+ 1/(t_{th} \sigma)^2} 
 \left( {1 \over t_{th} \sigma} \right)^2 \right], \label{eq:S1_I}\\
&& F2(r)=F1(r) (p/\rho g),\qquad \mathcal{F}2(r)=\mathcal{F}1(r) (p/ \rho g), \\
&& G2(r)=G1(r) (p/\rho g),\qquad \mathcal{G}2(r)=\mathcal{G}1(r) (p/ \rho g), \\
&& \mathcal{J}2(r)=\mathcal{J}1(r) (p/\rho g), \qquad \mathcal{H}2(r)=\mathcal{H}1(r) (p/\rho g), \\
&&\mathcal{S}2_R(r)=\mathcal{S}1_R(r) (p/ \rho g) , \qquad \mathcal{S}2_I(r)=\mathcal{S}1_I(r) (p/ \rho g)   \label{eq:S2_R} \\
&& \alpha_1(r)= - \rho\kappa_{th}{4\sigma_B  T^4 \over \pi}{1\over 1+ 1/(t_{th} \sigma)^2} {\Gamma_1 -1 \over \Gamma_1} 
=-{1\over 4 \pi}\sigma p {1\over t_{th} \sigma}{1\over 1+ 1/(t_{th} \sigma)^2} \\
&& \alpha_2(r)=  \rho\kappa_{th}{4\sigma_B  T^4 \over \pi} {1\over 1+ 1/(t_{th} \sigma)^2}{N^2 \over g}
={1\over 4 \pi}\sigma p {1\over t_{th} \sigma}{1\over 1+ 1/(t_{th} \sigma)^2}{N^2 \over g} {\Gamma_1 \over \Gamma_3-1}   \\
&& \alpha_3(r)= -\rho\kappa_{th}{4\sigma_B  T^4 \over \pi}
{1\over 1+ 1/(t_{th} \sigma)^2}
 {\Gamma_1-1 \over \Gamma_1}   {1\over t_{th} \sigma}=\alpha_1(r)  {1\over t_{th} \sigma} \\
&& \alpha_4(r)=  \rho\kappa_{th}{4\sigma_B  T^4 \over \pi} {1\over 1+1/(t_{th} \sigma)^2}
{N^2 \over g} {1\over t_{th} \sigma} = \alpha_2(r)  {1\over t_{th} \sigma}\\
&& \alpha_5(r)=\rho \kappa_{th} \left( 1-{1\over 1+ 1/(t_{th} \sigma)^2} \left( {1\over t_{th}\sigma} \right)^2  \right) \\
&& \alpha_6(r)= \rho \kappa_{th} {1\over 1+ 1/(t_{th} \sigma)^2} {1\over t_{th} \sigma}\\
&& \mathcal{S}3_R (r)= \rho \kappa_{th} \left[ {1 \over 4 \pi}{1\over 1+ 1/(t_{th} \sigma)^2} \left( {1\over t_{th} \sigma} \right)^2  \right] c_{m=2,n} \gamma F_* \exp(-p \kappa_*/g),\\
&& \alpha_7(r)= \rho \kappa_{th}{4\sigma_B T^4\over \pi}  
{1\over 1+ 1/(t_{th} \sigma)^2}
{\Gamma_1-1 \over \Gamma_1} 
{1 \over t_{th} \sigma} = -\alpha_3(r) \\
&& \alpha_8(r)= -\rho \kappa_{th}{4\sigma_B T^4\over \pi}  
{1\over 1+ 1/(t_{th} \sigma)^2}
{N^2 \over g} 
{1\over t_{th} \sigma} = -\alpha_4(r)\\
&& \alpha_9(r)=- \rho \kappa_{th}{4\sigma_B T^4\over \pi} 
{1\over 1+ 1/(t_{th} \sigma)^2}
{\Gamma_1 -1 \over \Gamma_1}  =\alpha_1(r)\\
&& \alpha_{10}(r)= \rho \kappa_{th}{4\sigma_B T^4\over \pi} 
{1\over 1+ 1/(t_{th} \sigma)^2}
 {N^2 \over g} =\alpha_2(r)\\
&& \alpha_{11}(r)= -\rho \kappa_{th}
{1\over 1+ 1/(t_{th} \sigma)^2}{1\over t_{th} \sigma}= -\alpha_6(r)\\
&& \alpha_{12}(r)=\rho \kappa_{th} \left( 1 -
{1\over 1+ 1/(t_{th} \sigma)^2}
\left( {1\over t_{th} \sigma} \right)^2 
\right)  = \alpha_5(r)\\
&& \mathcal{S}3_I(r) = \rho \kappa_{th} {1\over 4\pi}
{1\over 1+ 1/(t_{th} \sigma)^2}
{1 \over t_{th} \sigma} c_{m=2,n} \gamma F_* \exp(-p\kappa_*/g).\\
\end{eqnarray}
It can be easily observed from the above equations that when $F_*=0$, all the source terms $S1_R$, $S1_I$, $\mathcal{S}2_R$, $\mathcal{S}2_I$,  $\mathcal{S}3_R$, and  $\mathcal{S}3_I$ vanish, thereby leading to no perturbations. 

Using $T'/T=p'/p-\rho'/\rho$ and the energy equation, we can express $T'/T$ in terms of the perturbed variables to be solved; i.e.,
\begin{eqnarray}
&&{T'_R \over T'}={1\over 1+ ({\rho \over p} {\Gamma_3 -1 \over \Gamma_1} {1 \over \sigma} \sigma_B 16 T^4\kappa_{th})^2} \left\{ {\Gamma_1 -1 \over \Gamma_1} y_{p,R} - {N^2 \over g} \xi_{r,R}-{\rho \over p}{\Gamma_3 -1 \over \Gamma_1} {1 \over \sigma} 4 \pi \kappa_{th} J'_{th,I} \right. \label{eq:T'_R}\\
&& \left. +  \left[ {\Gamma_1 -1 \over \Gamma_1} y_{p,I} - {N^2 \over g} \xi_{r,I}+{\rho \over p}{\Gamma_3 -1 \over \Gamma_1} {1 \over \sigma} [c_{m=2,n}\kappa_* F_* \exp(-p\kappa_*/g)+4 \pi \kappa_{th} J'_{th,R} ] \right] {\rho \over p}{\Gamma_3 -1\over \Gamma_1} {1 \over \sigma} \sigma_B 16 T^4 \kappa_{th} \right\},\nonumber \\
&&{T'_I \over T'}={1\over 1+ ({\rho \over p} {\Gamma_3 -1 \over \Gamma_1} {1 \over \sigma} \sigma_B 16 T^4\kappa_{th})^2} \times \label{eq:T'_I} \\
&&\left\{ {\Gamma_1 -1 \over \Gamma_1} y_{p,I} - {N^2 \over g} \xi_{r,I}+{\rho \over p}{\Gamma_3 -1 \over \Gamma_1} {1 \over \sigma} [c_{m=2,n}\kappa_* F_* \exp(-p\kappa_*/g)+ 4 \pi \kappa_{th} J'_{th,R}] \right. \nonumber \\
&& \left. -  \left[ {\Gamma_1 -1 \over \Gamma_1} y_{p,R} - {N^2 \over g} \xi_{r,R}-{\rho \over p}{\Gamma_3 -1 \over \Gamma_1} {1 \over \sigma} 4 \pi \kappa_{th} J'_{th,I} \right] {\rho \over p}{\Gamma_3 -1\over \Gamma_1} {1 \over \sigma} \sigma_B 16 T^4 \kappa_{th}  \right\}.\nonumber
\end{eqnarray}
The expressions of $y_T$ are used to connect the $y_T$ from the planet interior at the interface.

\section{Linear equation for the interior} \label{sec:interior}
From eqns(\ref{eq:perb1}), (\ref{eq:perb2}), (\ref{eq:pert3}), and (\ref{eq:pert4}), the linear equations for the planet interior to solve for
$y_p$, $\xi_r$, $F'_r$, and $y_T\equiv T'/T$ read 
\begin{eqnarray}
&&{dy_{p,R} \over dr} + f_p(r)y_{p,R} +  f_{\xi_r}(r) \xi_{r,R} + f_T(r)y_{T,R} =0,\label{eq:MESAf}\\
&&{dy_{p,I} \over dr} + f_p(r)y_{p,I} +  f_{\xi_r}(r) \xi_{r,I} + f_T(r)y_{T,I} =0,\\
&&{d \xi_{r,R} \over dr} + g_p(r)y_{p,R} + g_{\xi_r}(r) \xi_{r,R} + g_T(r) y_{T,R} =0,\\
&&{d \xi_{r,I} \over dr} + g_p(r)y_{p,I} + g_{\xi_r}(r) \xi_{r,I} + g_T(r) y_{T,I} =0,\\
&&{d F'_{r,R} \over dr}  - {\rm Im}[h_p(r)] y_{p,I}
 -{\rm Im}[h_{\xi_r}(r)] \xi_{r,I}  + {\rm Re}[h_T(r)] y_{T,R} - {\rm Im}[h_T(r)] y_{T,I} \nonumber \\
&& \qquad +h_{F}(r) F'_{r,R}=0,\\
&&{d F'_{r,I} \over dr} + {\rm Im}[h_p(r)] y_{p,R}
+{\rm Im}[h_{\xi_r}(r)] \xi_{r,R} \nonumber \\
&& \qquad + {\rm Im}[h_T(r)] y_{T,R} + {\rm Re}[h_T(r)] y_{T,I}
+h_{F}(r) F'_{r,I}=0,\\
&&{d y_{T,R} \over dr}+ I_p(r) y_{p,R} + I_T(r) y_{T,R} + I_{F}(r) F'_{r,R} =0,\\
&&{d y_{T,I} \over dr}+ I_p(r) y_{p,I} + I_T(r) y_{T,I} + I_{F}(r) F'_{r,I} =0,\label{eq:MESAl}
\end{eqnarray}
where the coefficients are
\begin{eqnarray}
&&f_p = -{\rho g \over p}\left[ 1- \left( {\partial \ln \rho \over \partial \ln p } \right)_T  \right],\\
&&f_{\xi_r} = -{\sigma^2 \over p/\rho},\\
&&f_T = {\rho g \over p} \left( {\partial \ln \rho \over \partial \ln T} \right)_p, \\
&&g_p = \left( {\partial \ln \rho \over \partial \ln p} \right)_T  - {\lambda_n \over \sigma^2 r^2 }{p\over \rho},\\
&&g_{\xi_r} = {2\over r} + {d \ln \rho \over dr},\\
&&g_T =  \left( {\partial \ln \rho \over \partial \ln T} \right)_p \\
&&{\rm Im}[h_p] = { \sigma p \over \Gamma_3 -1} \left[  \Gamma_1 \left( {\partial \ln \rho \over \partial \ln p }  \right)_T -1 \right] 
=-{1\over \eta}F   \left[  \Gamma_1 \left( {\partial \ln \rho \over \partial \ln p }  \right)_T -1 \right], \label{eq:eta1}\\
&&{\rm Im}[h_{\xi_r}] = -{\sigma p \over \Gamma_3 -1} \Gamma_1 A = {1\over \eta}F \Gamma_1 A, \label{eq:eta2} \\
&& {\rm Re}[{h_T}]=- \lambda_n {\Lambda F \over r^2}, \\
&& {\rm Im}[{h_T}] =  {\sigma p \over \Gamma_3 -1} \Gamma_1  \left( { \partial \ln \rho \over \partial \ln T} \right)_p 
=- {1\over \eta}F\Gamma_1  \left( { \partial \ln \rho \over \partial \ln T} \right)_p ,\label{eq:eta3}\\
&&h_{F} = {2\over r},\\
&&I_p = -{1\over \Lambda} (1+\kappa_\rho) \left( {\partial \ln \rho \over \partial \ln p} \right)_T,\\
&&I_T = -{1\over \Lambda}  \left[ (1+\kappa_\rho ) \left( { \partial \ln \rho \over \partial \ln T }\right)_p - (4-\kappa_T) \right] \\
&&I_{F} = -{1\over F \Lambda},
\end{eqnarray}
and the definitions of $\Lambda$, $F \Lambda$, $\kappa_T$, and $\kappa_\rho$ are given by eqns(\ref{eq:L}), (\ref{eq:FL}) and (\ref{eq:dkappa}).
The coefficients are evaluated from the background states provided by MESA. When $\eta \rightarrow 0$ in eq(\ref{eq:pert3}), i.e. in eqns(\ref{eq:eta1}), (\ref{eq:eta2}), and (\ref{eq:eta3}), the above set of ODEs for the interior reduces to the ODEs (\ref{eq:perb1})-(\ref{eq:perb3}) in the absence of the non-adiabatic term.




\end{document}